\documentclass[10pt]{iopart}

\usepackage{setstack}

\usepackage{amssymb}
\usepackage{amsfonts}
\usepackage[dvips]{color}
\usepackage{epsfig}
\usepackage{iopams}

\def\schr{Schr\"{o}dinger }
\def\Ho{\hat{H}}

\def\I{\hat{\mathrm{I}}}
\def\Ko{\hat{K}}
\def\ket#1{|#1\rangle}
\def\kett#1{#1\rangle}
\def\bra#1{\langle#1|}

\def\d{\mathrm{d}}
\def\v#1{\mathbf{#1}}
\def\Tp{T_{\mathrm{P}}}
\newcommand{\kvs}{\hat{\mathbf{k}}}
\newcommand{\yv}{\mathbf{y}}
\newcommand{\Kvs}{\hat{\mathbf{K}}}
\def\Mp{M_{\mathrm{P}}}

\def\lr{L_{\mathrm{\scriptsize{R}}}}
\def\llr{\lambda_{\mathrm{\scriptsize{R}}}}
\def\kcut{k_\lambda}
\def\Pr#1{\ket{#1}\bra{#1}}
\def\tex{T_{\mathrm{\scriptsize{ex}}}}
\def\mc70{M_{C_{70}}}

\def\vg#1{\boldsymbol{#1}}
\def\a{\alpha}
\def\b{\beta}

\def\pd{\partial}
\def\abs#1{\left|#1\right|}
\def\Lp{L_{P}}
\def\m#1{\left\langle#1\right\rangle}
\newcommand{\xv}{\mathbf{x}}
\newcommand{\kv}{\mathbf{k}}
\def\O{\Omega}
\def\cd{\nabla}

\def\ms#1{\langle#1\rangle}

\def\ee{Einstein's equation }
\def\schr{Schr\"{o}dinger }
\def\lamm{L\"{a}mmerzahl }

\def\pl{\widetilde{p}}
\def\vl{\widetilde{v}}
\def\dv{\delta v}

\def\phit{\Phi}

\def\ex{\exp{\left(-iMc^2 t/\hbar\right)}}
\def\cI{C_1}
\def\cII{C_2}

\begin{document}

\title[Dephasing due to
conformal fluctuations]{Dephasing of a non-relativistic quantum
particle due to a conformally fluctuating spacetime}

\author{Paolo M. Bonifacio$^{1}$, Charles H.-T. Wang$^{1,2}$, J. Tito Mendon\c{c}a$^{2,3}$, Robert Bingham$^{2,4}$}
\address{$^1$SUPA Department of Physics,
University of Aberdeen, King's College, Aberdeen AB24 3UE, UK\\
$^2$Rutherford Appleton Laboratory, STFC, Chilton, Didcot,
Oxfordshire OX11 0QX, UK\\
$^3$CFP and CFIF, Instituto Superior T\'{e}cnico, 1049-001 Lisboa,
Portugal\\
$^4$SUPA Department of Physics, University of Strathclyde, Glasgow
G4 0NG, UK} \ead{\mailto{p.bonifacio@abdn.ac.uk},
\mailto{c.wang@abdn.ac.uk}, \mailto{titomend@ist.utl.pt},
\mailto{r.bingham@rl.ac.uk}}

\begin{abstract}
We investigate the dephasing suffered by a nonrelativistic quantum
particle within a conformally fluctuating spacetime geometry.
Starting from a minimally coupled massive Klein-Gordon field, the
low velocity limit yields an effective \schr equation where the wave
function couples to gravity through an effective nonlinear potential
induced by the conformal fluctuations. The quantum evolution is
studied through a Dyson expansion scheme up to second order. We show
that only the nonlinear part of the potential can induce dephasing.
This happens through an exponential decay of the off diagonal terms
of the particle density matrix. The bath of conformal radiation is
modeled in 3-dimensions and its statistical properties are described
in general in terms of a power spectral density. The case of a
Lorentz invariant spectral density, allowing to model vacuum
fluctuations at a low energy domain, is investigated and a general
formula describing the loss of coherence derived. This depends
quadratically on the particle mass and on the inverse cube of a
typical particle dependent cutoff scale. Finally, the possibilities
for experimental verification are discussed. It is shown that
current interferometry experiments cannot detect such an effect.
However this conclusion may improve by using high mass entangled
quantum states.
\end{abstract}



\section{Introduction}

It is generally agreed that the underlying quantum nature of gravity
implies that the spacetime structure close to the Planck scale
departs from that predicted by General Relativity. Unfortunately the
Quantum Gravity domain is still beyond modern particle accelerators
such as LHC. Nonetheless, finding experimental ways to test the
quantum structure of spacetime would be highly beneficial to the
theoretical developments of our fundamental theories of nature. In
this respect it is has been suggested that quantum gravity could
induce decoherence on a quantum particle through its underlying
Planck scale spacetime fluctuations
\cite{powerpercival00,bingham05,wang2006,camelia05,anastopoulos07,breuer08}.

As the sensitivity and performance of matter wave interferometers is
increasing \cite{hornberger03, hackermuller03,
hornberger04,hackermuller04,stibor05}, it is important to assess the
theoretical possibility of a future experimental detection of
intrinsic, spacetime induced decoherence. The closely related
dephasing effect due to a random bath of classical GWs (e.g. of
astrophysical origin) has been extensively studied e.g. in
\cite{reynaud04}. The problem of the decoherence induced by
spacetime fluctuations is difficult to study as a quantum gravity
theory is still missing. Notwithstanding promising progress, mainly
in loop quantum gravity and superstring theory \cite{kiefer05}, a
coherent and established quantum gravity theoretical framework is
still missing. Thus any theoretical attempt for a prediction of the
decoherence induced by spacetime fluctuations must exploit some
semiclassical framework. Such approaches typically represent the
spacetime metric close to the Planck scale by means of fluctuating
functions. These are usually supposed to mimic the vacuum quantum
property of spacetime down to some cutoff scale $\ell = \lambda
\Lp,$ where $\Lp$ is the Planck scale. The adimensional parameter
$\lambda$ marks the benchmark between the fully quantum regime and
the scale where the classical properties of spacetime start to
emerge \cite{powerpercival00}. The fact that classical fluctuating
fields can be used to reproduce various genuine quantum effects is
well known, e.g. from the work of Boyer \cite{boyer1969,boyer75} in
the case of the EM field or Frederick \cite{frederick76} in the case
of spacetime fluctuations. This is often exploited in the literature
in relation to problems involving the microscopic behavior of the
spacetime metric; e.g. a stochastic metric was employed in
\cite{moffat97,miller99} to study the problem of gravitational
collapse and big bang singularities, while in \cite{goklu08}
spacetime metric fluctuations were introduced and their ability to
induce a WEP violation studied.

A pioneering analysis of the problem of spacetime induced
decoherence has been proposed by Power \& Percival (PP in the
following) \cite{powerpercival00} in the case of a conformally
modulated Minkowski spacetime with conformal fluctuations traveling
along 1 space dimension. This was improved by Wang et al.
\cite{wang2006}, who extended upon PP work attempting to include the
effect of GWs. Conformal fluctuations are interesting as they are
mathematically easy to treat and offer a convenient way to build
`toy' models to assess some of the problem's features. They have an
important role in theoretical physics \cite{kastrup08} and are
sometimes invoked in the literature also in relation to universal
scalar fields \cite{nelson85,miller99} that can arise naturally in
some modified theories of gravity such as scalar-tensor theories
\cite{dicke62,wagoner70}.

Within a semiclassical approach that `replaces' the true quantum
environment by classical fluctuating fields we should properly speak
of \emph{dephasing} of the quantum particle rather than decoherence.
In a remarkable paper \cite{stern90} about quantum interference in
the presence of an environment, Stern and co-authors showed that a
fully quantum approach that studies decoherence by `tracing away'
the environment degrees of freedom in the quantum system made up by
system + environment, and that in which the dephasing of the quantum
particle is due to a stochastic background field give equivalent
results.

In this paper we consider a conformally modulated 4-dimensional
spacetime metric of the form $g_{ab} = (1 + A)^2 \eta_{ab}$. Such a
metric has been considered by PP \cite{powerpercival00}, where the
dephasing problem was studied in the simple idealized case of a
particle propagating in 1-dimension. By imposing \ee on the metric
$g_{ab}$, PP deduced a wave equation for $A$. Their procedure to
derive an effective newtonian potential interacting with the quantum
probe started from the geodesic equation of a test particle. Even
though this didn't take properly into account the nonlinearity in
the conformal factor $(1 + A)^2$, they found correctly that the
change in the density matrix is given by $\delta \rho \propto M^2 T
A_0^4 \tau_*$, where $M$ is the probing particle mass, $T$ the
flight time, $A_0^4$ the amplitude of the conformal fluctuations and
$\tau_*$ their correlation time. This formula was used to set limits
upon $\lambda$. However in doing this they did not treat the
statistical properties of the fluctuations properly and this
resulted in the wrong estimate $\lambda \propto (M^2 T / \delta
\rho)^{1/7}$, as already noted by Wang et al in \cite{wang2006}.

In their work, Wang and co-authors attempt to include GWs into the
analysis by considering a metric of the kind $g_{ab} = (1 + A)^2
\gamma_{ab}$. This was done by exploiting the results in
\cite{wang05a,wang05b} where a canonical geometrodynamics approach
employing a conformal spacial 3-metric was studied. By exploiting an
energy density balancing mechanism between the conformal and GWs
parts of the total gravitational Hamiltonian the statistical
properties of the conformal fluctuations where fixed. This
corresponded to assume that each `quantum' of the conformal field
possessed a zero point energy $-\hbar \omega$. Though an improvement
over PP work, this approach is still 1-dimensional and too crude to
make predictions. Moreover, as it shall be discussed extensively in
a future report \cite{bonifacio2009II}, the issue of energy balance
between conformal fluctuations and GWs is a delicate one, and likely
not to occur within the standard GR framework.

In the present work we provide a coherent 3-dimensional treatment of
the problem of a slow massive test particle coupled to a conformally
fluctuating spacetime. The conformal field $A$ is assumed to satisfy
a simple wave equation. This will allow a direct comparison with PP
result. We also notice that such a framework is expected to arise
naturally within a scalar-tensor theory of gravity. This issue will
be discussed in a future report \cite{bonifacio2009III}.

The work is organized as follows: in \emph{\sref{lvle}} the correct
non-relativistic limit of a minimally coupled Klein-Gordon field is
deduced and an effective newtonian potential depending nonlinearly
on $A$ is identified in the resulting effective \schr equation. In
\sref{aqe} we set the general formalism to study the average quantum
evolution through a Dyson expansion scheme for the particle density
matrix $\rho$. In \emph{\sref{cfcp}}, general results derived in
\ref{ap3} are used to model the statistical and correlation
properties of the fluctuations through a general, unspecified, power
spectral density. In \emph{\sref{sec1} and \ref{lemma}} we compute
the average quantum evolution and derive a general expression for
the evolved density matrix. We show in general that only a nonlinear
potential can induce dephasing. The resulting dephasing formula
implies an exponential decay of the density matrix off-diagonal
elements and is shown to hold in general and independently of the
specific spectral properties of the fluctuations.  All we assume is
that these obey a simple wave equation and that they are a zero mean
random process. The overall dephasing predicted within the present
3-dimensional model -equation \eref{P2_3d}- is seen to be about two
orders of magnitudes larger than in the 1-dimensional case as
derived by PP. Next we consider in \emph{\sref{pcf}} the problem of
vacuum fluctuations. To this end a power spectrum $S(\omega) \sim 1
/ \omega$ is introduced and we derive an explicit formula for the
rate of change of the density matrix. This result improves over both
Percival's and Wang's work in that its key ingredients are general
enough to be potentially suited for a variety of physical
situations. Finally the discussion in \emph{\sref{disc}} addresses
the question of whether the dephasing due to conformal vacuum
spacetime fluctuations could be detected. In other words, whether
the proposed theory can be falsified or not. A possibility would be
through matter wave interferometry employing large molecules. We
consider this issue in the final part of this paper by estimating
the probing particle resolution scale, setting its ability to be
affected by the fluctuations. The resulting formula for the
dephasing rate indicates that the level of the effect is still
likely to be beyond experimental capability, even for large
molecules (e.g. fullerenes) \cite{hackermuller04}. A measurable
effect could possibly result for larger masses, e.g. if entangles
quantum states were employed \cite{everitt08}.

\section{Low velocity limit and effective \schr
equation}\label{lvle}

The problem we wish to solve is clearly defined: we consider a
scalar field $A$ inducing conformal fluctuations on an otherwise
flat spacetime geometry according to
\begin{equation}\label{e1}
g_{ab} = (1 + A)^2 \eta_{ab},
\end{equation}
where $\eta_{ab} = \mathrm{diag}(-1,1,1,1,)$ is the Minkowski
tensor. We will refer to $A$ as to the \emph{conformal field} and
this will be assumed to satisfy the wave equation $\pd^c\pd_c A =
0$. Solving this equation with random boundary conditions results in
a randomly fluctuating field propagating in 3-dimensional space. We
assume this to be a small first order quantity, i.e. $\abs{A} =
O(\varepsilon \ll 1)$. \Eref{e1} expresses the spacetime metric in
the laboratory frame. We also suppose that the typical wavelengths
of $A$ are effectively cut off at a scale set by $\ell := \lambda
\Lp,$ where $\Lp = (\hbar G / c^3)^{1/2} \approx 10^{-35}$ m is the
Planck length. The adimensional parameter $\lambda$ represents a
structural property of spacetime marking the quantum-classical
transition: below $\ell$ a full quantum treatment of gravity would
be needed so that, by definition, $\ell$ represents the scale at
which a semiclassical approach that treats quantum effects by means
of classical randomly fluctuating fields is supposed to be a valid
approximation. The value for $\lambda$ is model dependent but it is
generally agreed that $\lambda \gtrsim 10^2$ \cite{wang2006}, so
that $\ell$ is expected to be extremely small from a macroscopic
point of view. This motivates the assumption that classical
macroscopic bodies, including the objects making up the laboratory
frame and also the observers, are unaffected by the fluctuations in
$A$. This corresponds to the idea that a physical object is
characterized by some typical resolution scale $\lr$ that sets its
ability to `feel' the fluctuations: if $\lr \gg \ell$ these average
out and do not affect the body, that simply follows the geodesic of
the flat background metric. On the other hand a microscopic particle
can represent a successful probe of the conformal fluctuations if
its resolution scale is small enough.

We are interested in the change of the phase induced on the wave
function of a quantum particle by the fluctuating gravitational
field. Various approaches to the problem of how spacetime curvature
affects the propagation of a quantum wave exist in the literature;
e.g. for a stationary, weak field and a non relativistic particle a
\schr-like equation can be recovered \cite{dewitt66}. The more
interesting case of time varying gravitational fields can be treated
e.g. by eikonal methods that are usually restricted to weak fields
with $g_{ab} = \eta_{ab} + h_{ab}$ and $\abs{h_{ab}} \ll 1$
\cite{linet76,cai89}. Other approaches, e.g. in
\cite{goklu08,breuer08}, are based on the scheme developed by Kiefer
\cite{kiefer91} for the nonrelativistic reduction of a Klein-Gordon
field which is minimally coupled to a linearly perturbed metric. The
approach of PP in \cite{powerpercival00} and of Wang at el. in
\cite{wang2006} was to derive the geodesic equation in the weak
field limit. Their treatments were however employing, incorrectly,
the usual newtonian limit scheme which is valid only for weak,
linear and static perturbations \cite{wald84}. Conceptually the wave
approach is more satisfying than that based on the geodesic equation
because, e.g., the coupling between gravity and a scalar field is
well understood and in the appropriate non-relativistic weak field
limit an effective \schr equation emerges. This will be our approach
below.

We describe the quantum particle of mass $M$ by means of a minimally
coupled Klein-Gordon (KG) field $\phi$:
\begin{equation*}
g^{ab}\cd_a\cd_b \phi = \frac{M^2 c^2}{\hbar^2} \phi,
\end{equation*}
where $\cd_a$ is the covariant derivative of the physical metric
$g_{ab}$. Using $g_{ab} = \O^2 \eta_{ab}$ this equation can easily
be made explicit \cite{wald84} and reads:
\begin{equation}\label{KG1}
\left( -\frac{1}{c^2}\frac{\pd^2}{\pd t^2} + \nabla^2 \right) \phi =
\frac{\Omega^2 M^2 c^2}{\hbar^2} \phi - 2 \pd_a (\ln \Omega) \pd^a
\phi,
\end{equation}
i.e. the wave equation for a massive scalar field plus a
perturbation due to $A$ describing the coupling to the conformally
fluctuating spacetime. We need to take the appropriate
non-relativistic limit in order to deduce an effective \schr
equation. Before doing this we remark that, had we consider the
alternative meaningful scenario of a conformally coupled scalar
field, then the equation $g^{ab}\cd_a\cd_b \phi - R\phi/6 - M^2 c^2
\phi / \hbar^2 = 0$ would read explicitly
\begin{equation*}
\left( -\frac{1}{c^2}\frac{\pd^2}{\pd t^2} + \nabla^2 \right) \phi =
\frac{\Omega^2 M^2 c^2}{\hbar^2} \phi - 2 \pd_a (\ln \Omega) \pd^a
\phi - \phi \O^{-1}\pd^c \pd_c \O.
\end{equation*}
Since it is $\O^{-1} \pd^c \pd_c \O = (1 - A) \pd^c \pd_c A +
O(\varepsilon^3)$ we see that \emph{if $A$ is assumed to satisfy the
wave equation then the curvature term has no effect}: in this case
the minimally and conformally coupled KG equations are equivalent up
to second order in $A$. We also note that, by introducing the
\emph{auxiliary field} $\phit := \O \phi$, equation \eref{KG1} turns
out to be equivalent to:
\begin{equation*}
\left( -\frac{1}{c^2}\frac{\pd^2}{\pd t^2} + \nabla^2 \right) \phit
= \frac{\Omega^2 M^2 c^2}{\hbar^2} \phit.
\end{equation*}
In principle, if a solution for $\phit$ were known, then \emph{the
physical scalar field representing the particle} would follow
formally as $\phi = \O^{-1}\phit = (1 - A + A^2) \phit,$ up to
second order in $A$. However in studying the dephasing problem we
will \emph{only} find an averaged solution for the average density
matrix representing the quantum particle. Therefore, even if a
solution in this sense is know in relation to $\phit$, it would not
be obvious how to obtain the corresponding averaged density matrix
related to $\phi$, which is what we are interested in.

In view of the above considerations we work directly with equation
\eref{KG1} and now proceed in deriving its suitable non-relativistic
limit. We will make two assumptions:
\begin{enumerate}
  \item the particle is slow i.e., if $\pl = M \vl$ is its momentum in the
  laboratory, we have:
  \begin{equation*}
  \frac{\vl}{c} \ll 1;
  \end{equation*}
  \item the effect of the conformal fluctuations is small, i.e. the
  induced change in momentum $\delta p = M \dv$ is small compared to
  $M \vl$:
  \begin{equation*}
  \frac{\dv}{\vl} \ll 1.
  \end{equation*}
\end{enumerate}
In view of these assumptions we can write:
\begin{equation*}
\phi = \psi \ex,
\end{equation*}
where the field $\psi$ is close to be a plane wave of momentum
$\pl$. As a consequence we have:
\begin{equation*}
\frac{\pd^2 \psi}{\pd t^2} \approx
-\frac{1}{\hbar^2}\left(\frac{\pl^2}{2M}\right)^2\psi.
\end{equation*}
Using this and multiplying by $\hbar^2 / 2M$, equation \eref{KG1}
yields:
\begin{eqnarray}\label{eqeqeq}
\fl\left[ \underbrace{i\hbar\frac{\pd}{\pd t} +
\frac{\hbar^2}{2M}\nabla^2}_{T_1}
-\underbrace{\frac{Mc^2}{8}\left(\frac{\vl}{c}\right)^4}_{T_2}
\right] \psi = \nonumber\\
\hspace{-1cm}=\underbrace{\left(A +
\frac{A^2}{2}\right)Mc^2}_{T_3}\psi -
\underbrace{\frac{\hbar^2}{M}\pd^a \phi\pd_a
\ln(1+A)\times\exp{\left(iMc^2 t/\hbar\right)}}_{T_4}.
\end{eqnarray}

Leaving the term $T_4$ aside for the moment, the orders of the three
underlined terms must be carefully assessed. We have:
\begin{equation*}
T_1 \sim M\vl^2,\quad T_2 \sim Mc^2\left(\frac{\vl}{c}\right)^4\quad
\m{T_3} \sim \varepsilon^2 M c^2,
\end{equation*}
where an average has been inserted since $T_3$ is fluctuating. It
follows that
\begin{equation*}
\frac{T_2}{T_1} \sim \left(\frac{\vl}{c}\right)^2 \ll 1.
\end{equation*}
Thus, in the non-relativistic limit, $T_2$ is negligible in
comparison to $T_1$. This is the case in a typical interferometry
experiment where it can be $\vl \approx 10^2$ m s$^{-1}$
\cite{stibor05}, so that $(\vl / c)^2 \sim 10^{-12}$. Next we have:
\begin{equation*}
\frac{\m{T_3}}{T_1} \sim \left(\frac{\varepsilon c}{\vl}\right)^2.
\end{equation*}
The request that the conformal fluctuations have a small effect thus
gives the condition
\begin{equation}\label{C2}
\frac{\m{T_3}}{T_1} \ll 1 \quad \Leftrightarrow \quad \varepsilon^2
\sim \m{A^2} \ll \left(\frac{\vl}{c}\right)^2.
\end{equation}
That this condition is effectively satisfied can be checked a
posteriori after the model is complete. It depends on the
statistical properties of the conformal field and the particle
ability to probe them. This will be related to a particle
\emph{resolution scale}. At the end of the discussion in \sref{sneo}
we will show that \eref{C2} is satisfied if, e.g., the particle
resolution scale is given by its Compton length.

Under these conditions the non-relativistic limit of equation
\eref{eqeqeq} yields:
\begin{equation}\label{eqeq}
-\frac{\hbar^2}{2M}\nabla^2\psi + \left(A + \frac{A^2}{2}\right)Mc^2
\psi + T_4 = i\hbar\frac{\pd \psi}{\pd t},
\end{equation}
where
\begin{equation*}
T_4 := -\frac{\hbar^2}{M}\pd^a \phi\pd_a \ln(1+A) \times
\exp{\left(iMc^2 t/\hbar\right)}.
\end{equation*}
In order to assess the correction due to this term we split it into
two contributions by writing separately the time and space
derivatives. Using the fact that
\begin{equation*}
\frac{\pd \phi}{\pd t} \approx -\frac{i M c^2}{\hbar}\psi\ex
\end{equation*}
it is easy to see that:
\begin{equation}\label{t4}
T_4 = -i\hbar\left( \dot{A} - A\dot{A} \right)\psi - i\hbar \vl
\left( A_{,x} - A A_{,x} \right)\psi,
\end{equation}
where $\dot{A} := \pd A / \pd t$, $A_{,x} := \pd A / \pd x$ and
where we assumed that the particle velocity in the laboratory is
along the $x$ axis. In \ref{apx} we show that \emph{if $A$ is (i) a
stochastic isotropic perturbation and (ii) effectively fast varying
over a typical length $ \lambda_A = \kappa h / (Mc)$ related to the
particle resolution scale}, then $T_4$ reduces to:
\begin{equation}\label{t4bis}
T_4 = \left( - A + A^2 \right) \frac{Mc^2}{\kappa}\psi.
\end{equation}
Here $\kappa \sim 1$ is dimensionless and its precise value is
unimportant. The important point is that $T_4$ yields a
\emph{positive} extra nonlinear term in $A$ that adds up to what we
already have in \eref{eqeq}. Finally we get the \emph{effective
\schr equation}
\begin{equation*}
-\frac{\hbar^2}{2M}\nabla^2\psi + V \psi = i\hbar\frac{\pd \psi}{\pd
t},
\end{equation*}
where the \emph{nonlinear fluctuating potential} $V$ is defined by
\begin{equation}\label{pot}
V := \left(\cI A + \cII A^2\right)Mc^2.
\end{equation}
The values of the constants $\cI$ and $\cII$ depend on $\kappa$ and
$\cII$ \emph{is always strictly positive}. For $\kappa = 1$ it would
be $\cI = 0$ and $\cII = 3/2.$ For generality we will leave them
unspecified in the following treatment and consider $\kappa$ as a
constant of order one.

\section{Average quantum evolution}\label{aqe}

\subsection{Dyson expansion for short evolution
time}\label{P2_ipde}

We now have a rather well defined problem: that of the dynamics of a
non relativistic quantum particle under the influence of the
nonlinear potential \eref{pot}. The \schr equation describing the
dynamics of a free particle is suitable to describe the interference
patterns that could result e.g. in an interferometry experiment
employing cold molecular beams. When the particle in the beam
propagates through an environment, we are dealing with an open
quantum system. This in general suffers decoherence, resulting in a
loss of visibility in the fringes pattern
\cite{hornberger04,stibor05}. This is a well defined macroscopic
quantity. In the present semiclassical treatment the environment due
to spacetime fluctuations is represented, down to the semiclassical
scale $\ell$, by a sea of random radiation encoded in $A$ and
resulting in the fluctuating potential $V$. An estimate of the
overall dephasing can be obtained by considering the statistical
averaged dynamics of a single quantum particle interacting with $V$.
In practise we will need (i) to solve for the dynamics of a single
particle of mass $M$ and (ii) calculate the averaged wavefunction by
averaging over the fluctuations. The outcome of (i) would be some
sort of `fluctuating' wavefunction carrying, beyond the information
related to the innate quantum behavior of the system, that related
to the fluctuations in the potential. The outcome of (ii) is to
yield a general statistical result describing what would be obtained
in an experiment where many identical particles propagate through
the same fluctuating potential.

We thus consider the Hamiltonian operator $\Ho(t)=\Ho^0+\Ho^1(t),$
where $\Ho^0$ is the kinetic part while
\begin{equation*}
\Ho^1(t)=\int \d^3 x V(\xv,t) \ket{\xv}\bra{\xv},
\end{equation*}
is the perturbation due to the fluctuating potential energy. Here
$\ket{\xv}\bra{\xv}$ is the projection operator on the space spanned
by the position operator eigenstate $\ket{\xv}$. Indicating the
state vector at time $t$ with $\psi_t$, the related \schr equation
reads
\begin{equation*}
\Ho(t)\psi_t = i\hbar\frac{\pd\psi_t}{\pd t}.
\end{equation*}
Using the density matrix formalism, the general solution can be
expressed through a Dyson series as \cite{roman65}
\begin{equation*}
\fl\rho_T=\rho_0 + \Ko_1(T)\rho_0+\rho_0
\Ko_1^{\dag}(T)+\Ko_2(T)\rho_0+\Ko_1(T)\rho_0 \Ko_1^{\dag}(T)+\rho_0
\Ko_2^{\dag}(T)+\ldots,
\end{equation*}
where $\rho_0$ is the initial density matrix and the propagators
$\Ko_1(T)$ and $\Ko_2(T)$ are given by
\begin{equation*}
\Ko_1(T):=-\frac{i}{\hbar}\int_{0}^{T}{\Ho}(t'){\d}t',
\end{equation*}
\begin{equation*}
\Ko_2(T):=-\frac{1}{\hbar^2}
\int_{0}^{T}{\d}t'\int_{0}^{t'}{\d}t''{\Ho}(t'){\Ho}(t'').
\end{equation*}
In truncating the series to second order we assume that the system
evolves for a time $T$ such that $ T\ll T^*$, where $T^*$ is defined
as the typical time scale required to have a significant change in
the density matrix $\rho$.

The effect of the environment upon a large collection of identically
prepared systems is found by taking the average over the fluctuating
potential as explained above. Formally and up to second order we
have
\begin{equation*}
\fl\m{\rho_T}=\m{\rho_0 + \Ko_1(T)\rho_0+\rho_0
\Ko_1^{\dag}(T)+\Ko_2(T)\rho_0+\Ko_1(T)\rho_0 \Ko_1^{\dag}(T)+\rho_0
\Ko_2^{\dag}(T)}.
\end{equation*}
The average density matrix $\m{\rho_T}$ will describe the average
evolution of the system including the effect of dephasing.

It is straightforward to show that, up to second order in the
Dyson's expansion, the kinetic and potential parts of the
hamiltonian give independent, additive contributions to the average
evolution of the density matrix, i.e. $\m{\rho_T}=[\rho_T]_0 +
\m{[\rho_T]_1},$ where
\begin{eqnarray*}
\fl[\rho_T]_0:=\rho_0 + [\Ko_1(T)]_0\rho_0+\rho_0
[\Ko_1(T)]_0^{\dag}\nonumber\\
+[\Ko_2(T)]_0\rho_0+[\Ko_1(T)]_0\rho_0
[\Ko_1(T)]_0^{\dag}+\rho_0 [\Ko_2(T)]_0^{\dag},
\end{eqnarray*}
\begin{eqnarray}\label{P2_finale1}
\fl\m{[\rho_T]_1}:=\left\langle\rho_0 + [\Ko_1(T)]_1\rho_0+\rho_0
[\Ko_1(T)]_1^{\dag}\right.\nonumber\\
\left.+[\Ko_2(T)]_1\rho_0+[\Ko_1(T)]_1\rho_0
[\Ko_1(T)]_1^{\dag}+\rho_0 [\Ko_2(T)]_1^{\dag}\right\rangle.
\end{eqnarray}
Here the kinetic propagators $[\Ko_1(T)]_0$ and $[\Ko_2(T)]_0$
depend solely on $\Ho^0$, while the potential propagators
$[\Ko_1(T)]_1$ and $[\Ko_2(T)]_1$ depend only on $\Ho^1(t)$. In the
next section we estimate the dephasing by calculating the term
$\m{[\rho_T]_1}$ alone.

\section{The conformal field and its correlation
properties}\label{cfcp}

We now set the statistical properties of the conformal field $A$.
This is assumed to represent a real, stochastic process having a
zero mean. We further assume it to be isotropic. In \ref{ap3} we
review a series of important results concerning stochastic
processes, in particular in relation to real stochastic signals
satisfying the wave equation. The main quantity characterizing the
process is the power spectral density $S(\omega)$. In the case of an
isotropic bath of random radiation, field averages such as
$\ms{A^2}$, $\ms{\abs{\nabla A}^2}$ and $\ms{(\pd_t A)^2}$ can be
found in terms of $S(\omega)$, e.g.
\begin{equation*}
\m{ A(\xv,t)^2} = \frac1{(2\pi)^3} \int \d^3 k\, S(k),
\end{equation*}
where $kc = \omega$. In \ref{ap3} we show how the conformal field
can be resolved into components traveling along all possible space
directions according to
\begin{equation*}
A(\v{x},t)=\int
\d\hat{\v{k}}\,A_{\hat{\v{k}}}(\hat{\v{k}}\cdot\v{x}/c-t),
\end{equation*}
where $\d\hat{\v{k}}$ indicates the elementary solid angle. The
capacity of the fluctuations to maintain correlation is encoded in
the autocorrelation function $C(\tau)$. In the same appendix we
prove a generalization of the usual Wiener-Khintchine theorem, valid
for the case of a spacetime dependent process satisfying the wave
equation, and linking the autocorrelation function to the Fourier
transform of the power spectral density according to:
\begin{equation}\label{WK}
C(\tau) := \frac{1}{(2\pi c)^3}\int\!\d \omega\, \omega^2
S(\omega)\cos(\omega\tau).
\end{equation}
This allows to prove that wave components traveling along
independent space directions are uncorrelated, i.e.
\begin{equation}\label{P2_cfc}
\m{ A_{\hat{\v{k}}}(t)\, A_{\hat{\v{k}}'}(t+\tau) }=
\delta(\v{k},\v{k}')\,C(\tau).
\end{equation}
The field mean squared amplitude is related to the correlation
function according to $\m{ A^2}=4\pi C_0$, as derived in \ref{ap3}

Isotropy implies that all directional components have the same
amplitude $A_0$. This is found introducing the \emph{normalized
correlation function} $R(\tau)$ through
\begin{equation*}
    R(\tau) := \frac{C(\tau)}{C(0)},
\end{equation*}
so that $R(0)=1$. Equation \eref{P2_cfc} can now be re-written as
$\m{ A_{\hat{\v{k}}}(t)\, A_{\hat{\v{k}}'}(t+\tau)}=
\delta(\v{k},\v{k}')\,C_0\,R(\tau)$ so that, introducing the
\emph{normalized directional components}, $ f_{\kvs}(t) :=
A_{\hat{\v{k}}}(t) / \sqrt{C_0} $ we have $\m{ f_{\hat{\v{k}}}(t)\,
f_{\hat{\v{k}}'}(t+\tau) }= \delta(\v{k},\v{k}')\,R(\tau)$. We now
define the constant $ A_0 := \sqrt{C_0} $ which is connected to the
\emph{squared amplitude per solid angle} according to $A_0^2 = C_0 =
\m{A^2} / 4\pi. $ The directional components are given by
$A_{\kvs}(t)=A_0\,f_{\kvs}(t)$ and the general conformal field can
finally be expressed as an elementary superposition of the kind
\begin{equation}\label{e6}
A(\xv,t)=A_0\int\! \d\kvs \,f_{\kvs}(\hat{\v{k}}\cdot\v{x}/c-t).
\end{equation}

\subsection{Summary of the correlation
properties of the conformal fluctuations}\label{P2_prop}

The main statistical properties of the directional stochastic waves
$f_{\kvs}$ are summarized by
\begin{equation}\label{e15}
\m{ f_{\kvs}(t) }=0,
\end{equation}
\begin{equation}\label{e17}
\m{ f_{\kvs}(t)f_{\kvs'}(t') }=\delta(\kvs,\kvs')R(t-t'),
\end{equation}
i.e. each component has \emph{zero mean} and fluctuations traveling
along different space directions are perfectly \emph{uncorrelated}.
These two properties imply that odd products of directional
components have also a zero mean, i.e.
\begin{equation}\label{e20}
\m{ f_{\kvs_1}(t_1)f_{\kvs_2}(t_2)f_{\kvs_3}(t_3) }=0.
\end{equation}
In the following dephasing calculation we will need to evaluate
means involving products of four directional components. To this
purpose we need to introduce the \emph{second order correlation
function} $R''(t-t')$ according to
\begin{equation}\label{e21}
\m{ [f_{\kvs}(t)]^2[f_{\kvs'}(t')]^2
}=1+\delta(\kvs,\kvs')[R''(t-t')-1].
\end{equation}
This definition is compatible with the fact that the mean is one
when components traveling in different direction are involved, i.e.
$\m{ [f_{\kvs}(t)]^2[f_{\kvs'}(t')]^2 } =1$ if $\kvs \neq \kvs'.$

\section{Dephasing calculation outline}\label{sec1}

To calculate the dephasing suffered by the probing particle we must
evaluate the average of all the individual terms in equation
\eref{P2_finale1}. The relevant propagators are
\begin{equation*}
[\Ko_1(T)]_1:= -\frac{i}{\hbar}\int_{0}^{T}\!\!\!{\d}t'{\Ho^1}(t'),
\end{equation*}
\begin{equation*}
[\Ko_2(T)]_1 := -\frac{1}{\hbar^2} \int_{0}^{T}\!\!\!\d
t\int_{0}^{t}{\d}t'{\Ho^1}(t){\Ho^1}(t').
\end{equation*}
The interaction Hamiltonian is given by
\begin{equation*}
\Ho^1(t)=\int \d^3 x V(\xv,t)\ket{\xv}\bra{\xv},
\end{equation*}
where the potential energy is
\begin{equation*}
    \fl V(\xv,t)=\cI Mc^2 A_0\int\! \d\kvs
    \,f_{\kvs}(t-\xv\cdot\kvs/c) + \cII M c^2A_0^2 \left[\int\! \d\kvs
    \,f_{\kvs}(t-\xv\cdot\kvs/c)\right]^2.
\end{equation*}

\subsection{First order terms of the Dyson expansion}

We evaluate the two first order terms in the Dyson expansion. For a
more compact notation, we do not show the argument of the
directional components $f_{\kvs}$. The contribution of the linear
part of the potential $\cI Mc^2 A$ vanishes trivially since $\m{A} =
0.$ The quadratic part gives:
\begin{equation*}
\fl\m{\Ko_1(T)\rho_0} = -\frac{i \cII M
c^2A_0^2}{\hbar}\int_{0}^{T}\!\!{\d}t\! \int\!\! \d^3 x\,
\ket{\xv}\bra{\xv}  \m{ \left[\int\! \d\kvs \,f_{\kvs}\right]^2 }
\rho_0.
\end{equation*}
Using $A_{\hat{\v{k}}}(t) = \sqrt{C_0} f_{\kvs}(t)$ and $\m{A^2} =
4\pi C_0 \equiv 4\pi A_0^2 $ it is seen that the average yields
$4\pi.$ Since $\int\!\! \d^3 x\, \ket{\xv}\bra{\xv} = \I$ and
integrating over $T$ we find
\begin{equation}\label{P2_1st}
\m{\Ko_1(T)\rho_0} = -\frac{4\pi \cII i M c^2A_0^2 T}{\hbar}\rho_0.
\end{equation}

The calculation of the other first order term proceeds in the same
way. Since $\Ko_1^{\dag}(T) = -\Ko_1$, it yields the same result as
in \eref{P2_1st} but with the opposite sign (more in general, all
the odd terms in the Dyson expansion have an $i$ factor and also
yield a vanishing contribution). We thus see that \emph{at first
order in the Dyson expansion there is no net dephasing and
$\ms{\Ko_1(T)\rho_0 + \rho_0\Ko_1^{\dag}(T)} = 0.$}

\subsection{Second order terms of the Dyson expansion}

The second order calculation is more complicated. A fundamental
point is that the linear part of the potential does again give a
vanishing contribution. Dephasing will be shown to come as a purely
nonlinear effect due to the nonlinear potential term $\sim A^2$.

\subsubsection{(Non)-contribution of the linear part of the potential}

To have an idea of how things work we consider e.g. the average of
the term $\Ko_2 \rho_0$. This has the following structure:
\begin{equation*}
\fl\m{\Ko_2 \rho_0} \sim \int {\d}t\int {\d}t' \int \d^3 y
\Pr{\yv}\int \d^3 y'\Pr{\yv'} \m{V(\yv,t)V(\yv',t')}\rho_0.
\end{equation*}
The interesting part is the average $\m{V(\yv,t)V(\yv',t')}$. This
is:
\begin{equation*}
\fl\m{V(\yv,t)V(\yv',t')} \sim \m{A(\yv,t)A(\yv',t')} +
\m{A(\yv,t)A^2(\yv',t')} + \m{A^2(\yv,t)A^2(\yv',t')}.
\end{equation*}
The first two term are due to the linear part of the potential. The
second of them vanishes in virtue of property \eref{e20}. This is
seen using the directional decomposition \eref{e6} and writing:
\begin{equation*}
\m{A(\yv,t)A^2(\yv',t')} = A_0^3 \int\! \d\kvs_1 \int\! \d\kvs_2
\int\! \d\kvs_3 \,\m{f_{\kvs_1}f_{\kvs_2}f_{\kvs_3}} = 0.
\end{equation*}
The first term results in the contribution:
\begin{eqnarray*}
\fl\m{A(\yv,t)A(\yv',t')} \Rightarrow \\
\fl\hspace{0.5cm}\int_0^T\!\!\! {\d}t\!\int_0^t \!\!\!{\d}t'\!\!
\int \d^3 y \Pr{\yv}\!\int \d^3 y'\Pr{\yv'}\! \int\! \d\kvs\!\!
\int\! \d\kvs'\m{f_{\kvs}(t - \yv\cdot\kvs)f_{\kvs'}(t' -
\yv'\cdot\kvs')}\rho_0.
\end{eqnarray*}
For convenience of notation we set $c=1$ in the arguments of the
directional functions $f_{\kvs}$. Using equation \eref{e17} the
average yields the 2-point correlation function according to
$\delta(\kvs,\kvs')R(t-t' + \yv'\cdot\kvs' - \yv\cdot\kvs)$.
Integrating with respect to $\kvs'$ yields:
\begin{eqnarray*}
\fl\m{A(\yv,t)A(\yv',t')} \Rightarrow \\
\int_0^T\!\!\!
{\d}t\!\int_0^t \!\!\!{\d}t'\!\! \int \d^3 y \Pr{\yv}\!\int \d^3
y'\Pr{\yv'}\! \int\! \d\kvs R[t-t' + \kvs\cdot(\yv' - \yv)] \rho_0.
\end{eqnarray*}
The corresponding matrix element is found by inserting $\bra{\xv}$
and $\ket{\xv'}$ respectively on the left and on the right. Using
$\bra{\xv}\kett{\yv} = \delta(\xv-\yv)$ and exploiting the
properties of the delta function we find
\begin{equation}\label{cont1}
\m{A(\yv,t)A(\yv',t')} \Rightarrow K \times \int_0^T\!\!\!
{\d}t\!\int_0^t \!\!\!{\d}t'\!\! \int\! \d\kvs R(t-t'),
\end{equation}
where $K$ is a constant given by
\begin{equation*}
K = -\frac{\cI^2 A_0^2 M^2 c^4\rho_{\xv\xv'}(0)}{\hbar^2},
\end{equation*}
and where $\rho_{\xv\xv'}(0) := \bra{\xv}\rho_0 \ket{\xv'}$. The
similar terms coming from $\m{\rho_0 \Ko_2^{\dag}}$ will contribute
in the same way as in \eref{cont1}, thus yielding an extra factor 2.
Finally, through a similar calculation it is found that the terms
$\sim A(\yv,t)A(\yv',t')$ coming from $\m{\Ko_1 \rho_0
\Ko_1^{\dag}}$ contribute according to:
\begin{equation*}
\fl\m{A(\yv,t)A(\yv',t')} \Rightarrow -K \times \int_0^T\!\!\!
{\d}t\!\int_0^T \!\!\!{\d}t'\!\! \int\! \d\kvs R[t-t' + \kvs\cdot
(\xv' - \xv)].
\end{equation*}
Bringing all together, the overall contribution deriving from the
linear part $\cI Mc^2 A$ of the effective potential is found to be
proportional to the expression:
\begin{equation}\label{I}
    I:=  \int_0^T \! {\d}t\left\{
2\int_0^t \!\!\!{\d}t' R(t-t') - \int_0^T \!\!\!{\d}t' R[t-t' +
\kvs\cdot (\xv' - \xv)]\right\}.
\end{equation}
In \ref{ap4} we prove that this vanishes provided $R(\tau)$ is an
even function and provided that the drift time $T$ is much larger
than the time needed by the fluctuations to propagate through the
distance $\abs{\Delta\xv}$, i.e. if $T \gg \kvs\cdot\Delta\xv$,
where $c=1$. This condition is certainly satisfied in a typical
interferometry experiment where the drift time $T$ can be of the
order of $\sim 1$ ms and $cT$ is indeed much larger than the typical
space separations $\abs{\Delta\xv}$ relevant to quantify the loss of
contrast in the measured interference pattern.

\emph{Thus we have here the important result that the linear part of
the potential doesn't induce in general any dephasing up to second
order in Dyson expansion}. In fact we show in the next section that
dephasing results \emph{purely} as an effect of the nonlinear
potential term $\cII M c^2 A^2.$

\subsubsection{Contribution of the nonlinear part of the potential}

This calculation requires estimating averages of the kind
$\m{A^2(\yv,t)A^2(\yv',t')}$, which will bring in the second order
correlation function $R''$ defined in \eref{e21}. This is
straightforward but algebraically lengthy. Proceeding in a similar
way as done above, exploiting the statistical properties
\eref{e15}-\eref{e21} and the already mentioned result $I = 0$ in
relation to \eref{I}, then the general result for the density matrix
and valid up to second order in the Dyson expansion can be proved to
be:
\begin{eqnarray}\label{P2_mainB}\nonumber
\fl\rho_{\xv\xv'}(T)=\,\,\rho_{\xv\xv'}(0) -\frac{32\cII\pi^2 M^2
c^4 A_0^4 \rho_{\xv\xv'}(0)}{\hbar^2} \times \left[
\int_{0}^{T}\!\!\!\d t \int_{0}^{T}\!\!\!{\d}t' R^2( t-t')
\right.\\\nonumber
\\
\left. \hspace{-2cm}- \frac{1}{16\pi^2} \int\! \d\kvs\int\! \d\Kvs\,
\int_{0}^{T}\!\!\!\d t \int_{0}^{T}\!\!\!{\d}t' R(
t-t'-\kvs\cdot\Delta\xv/c) \times R( t-t'-\Kvs\cdot\Delta\xv/c)
\right].
\end{eqnarray}

Remarkably, \emph{the second order correlation function doesn't play
any role: the first order correlation function $R(\tau)$ and thus
the power spectral density $S(\omega)$ completely determine the
system evolution up to second order.} Equation \eref{P2_mainB}
implies that the diagonal elements of the density matrix are left
unchanged by time evolution. This is seen by setting $\Delta \xv=0$
which yields immediately $\rho_{\xv\xv}(T)=\rho_{\xv\xv}(0)$ for
every $T$.

\section{General density matrix evolution for large drift
times}\label{lemma}

To verify that we have dephasing with an exponential decay of the
off diagonal elements we need further simplify the result
\eref{P2_mainB} by analyzing its behavior for appropriately large
evolution times. To this end we start from the following identity
\begin{eqnarray*}
\int_{0}^{T}\!\!\!\d t \int_{0}^{T}\!\!\!{\d}t' g(t-t') &=
\frac{1}{2\pi} \int_{0}^{T}\!\!\!\d t \int_{0}^{T}\!\!\!{\d}t'
\int_{-\infty}^{\infty}\!\!\!\d\omega\, \tilde{g}(\omega)e^{i\omega
(t-t')}
\nonumber\\
\nonumber\\
&= \frac{1}{2\pi}\int_{-\infty}^{\infty}\!\!\!\d\omega\,
\tilde{g}(\omega)\left[\frac{\sin(\omega T/2)}{\omega/2}\right]^2,
\label{TTtt00}
\end{eqnarray*}
where $\tilde{g}(\omega)$ denotes the Fourier transform of the
function $g(t)$. Denoting $[0,\Delta\omega]$ as a frequency interval
where $\tilde{g}(\omega)$ is slow varying, it is straightforward to
show that the above identity reduces to
\begin{equation}
\int_{0}^{T}\!\!\!\d t \int_{0}^{T}\!\!\!{\d}t' g(t-t') \approx
\tilde{g}(0) T, \label{TTtt01}
\end{equation}
for $T \gtrsim (\Delta\omega)^{-1}$. Note that for this to happen
$g(t)$ doesn't even need being an even function. This condition
translates what we mean by appropriately long evolution time. In
\sref{longdrift} we will show that it is equivalent to $T \gtrsim
\tau_*$, where $\tau_*$ is the fluctuations correlation time. This
is defined below.

Equation \eref{TTtt01} can now be used to evaluate the time
integrals appearing in \eref{P2_mainB}. This is done by identifying
in one case $g(t):=R^2(t)$ and in the other
$g_{\tau\tau'}(t):=R(t+\tau)R(t+\tau')$, where $\tau$ and $\tau'$
stand respectively for $-\kvs\cdot\Delta\xv$ and
$-\Kvs\cdot\Delta\xv$, and where the normalized correlation function
can be expressed, using the generalized W-K theorem \eref{WK}, as
\begin{equation*}
R(\tau) \equiv \frac{C(\tau)}{C_0} = \frac{1}{C_0(2\pi
c)^3}\int_0^{\omega_c}\!\!\!\d \omega\, \omega^2
S(\omega)\cos(\omega\tau).
\end{equation*}
Notice that the integration frequency has a cutoff at $\omega_c =
\omega_P / \lambda$, where the Planck frequency is $\omega_P := 2\pi
/ \Tp=1.166\times 10^{44}\mathrm{~s}^{-1}$. This is consistent with
the fact that below the scale $\ell = \lambda \Lp$ the approximation
of randomly fluctuating fields breaks down. In alternative this may
simply correspond to the fact that the probing particle is
insensitive to the short wavelengths as a result of its own finite
resolution scale $\lr.$

Application of \eref{TTtt01} to $g(t):=R^2(t)$ yields the result
\begin{equation}
\int_{0}^{T}\!\!\!\d t \int_{0}^{T}\!\!\!{\d}t' R^2(t-t') = \tau_{*}
T, \label{I1c}
\end{equation}
where the \emph{correlation time} is defined as
\begin{equation}\label{tts}
    \tau_{*} := \mathfrak{F}\left[R^2(t)\right](0) = \pi \frac{\int_0^{\omega_c}\d
\omega \,\omega^4\, S^2(\omega)}{\left[\int_0^{\omega_c}\d \omega
\,\omega^2\, S(\omega)\right]^2},
\end{equation}
$\mathfrak{F}$ denoting Fourier transform. On the other hand,
application of \eref{TTtt01} to
$g_{\tau\tau'}(t):=R(t+\tau)R(t+\tau')$ gives the result
\begin{equation}
\int_{0}^{T}\!\!\!\d t \int_{0}^{T}\!\!\!{\d}t' R(t+\tau)R(t+\tau')
= \tau_{*} \Gamma[\omega_c(\tau-\tau')]\, T, \label{I2c}
\end{equation}
where the \emph{characteristic function} $\Gamma$ has been defined
as
\begin{equation}\label{gamma}
\Gamma(\omega_c  t):=\frac{\int_0^{\omega_c}\d \omega\,\omega^4
S^2(\omega) \cos (\omega  t)}{\int_0^{\omega_c}\d \omega\,\omega^4
S^2(\omega)}.
\end{equation}
This is dimensionless and satisfies in general the following
properties:
\begin{itemize}
  \item $\Gamma(\omega_c  t)=\Gamma(-\omega_c  t)$,
  \item
  $\Gamma(0)=1,$
  \item $
\Gamma(\omega_c  t)<1,\quad\mathrm{for}\quad t\neq 0, $
  \item $\Gamma(\omega_c  t) \rightarrow 0, \quad \mathrm{for} \quad t
\rightarrow \infty.$
\end{itemize}
Notice that both the correlation time $\tau_{*}$ and the
characteristic function $\Gamma$ solely depend on the fluctuations
power spectral density.

The results \eref{I1c} and \eref{I2c} can now be used in equation
\eref{P2_mainB} to yield the neat result
\begin{equation}\label{afe2BBIISS}
\rho_{\xv\xv'}(T) = \rho_{\xv\xv'}(0)\left[1 - \frac{32\cII\pi^2 M^2
c^4 A_0^4\tau_{*} T}{\hbar^2} \times F(\Delta\xv) \right],
\end{equation}
where
\begin{equation}\label{fff}
F(\Delta\xv) := 1 - \frac{1}{16\pi^2} \int\! \d\kvs\! \int\!
\d\Kvs\, \Gamma[\omega_c(\Kvs-\kvs)\cdot\Delta\xv/c].
\end{equation}
This equation is important and represents one of the main result of
this paper. It implies that \emph{dephasing due to conformal
fluctuations does indeed occur in general and independently of the
precise power spectrum characterizing the fluctuations}. Without the
need to evaluate the angular integrals, this follows from the
properties of the characteristic function $\Gamma$. The fact that
$\Gamma[\omega_c t]<1$ implies $0 \leq F(\Delta\xv) \leq 1$ with (i)
$F(\Delta\xv = 0) = 0$ and (ii) $F(\Delta\xv \rightarrow \infty) =
1$ as special limiting cases. As a consequence the diagonal elements
are unaffected while the off-diagonal elements decay exponentially
according to
\begin{equation*}
\fl\dot{\rho}_{\xv\xv'}(0) := \frac{\rho_{\xv\xv'}(T) -
\rho_{\xv\xv'}(0)}{T} =  - \left[\frac{32\cII\pi^2 M^2 c^4
A_0^4\tau_{*}}{\hbar^2} \times F(\Delta\xv)\right]
\rho_{\xv\xv'}(0),
\end{equation*}
providing of course that $T$ is small enough so that the change in
the density matrix is small. Finally, if $\delta\rho :=
\rho_{\xv\xv'}(T) - \rho_{\xv\xv'}(0)$, we can define the
\emph{dephasing rate} as $\abs{\delta\rho/\rho_0}$. Thanks to the
property $F(\Delta\xv \rightarrow \infty) = 1$, this converges for
large spacial separations to the constant maximum value
\begin{equation}\label{P2_3d}
\abs{\frac{\delta\rho}{\rho_0}} =  \frac{32\cII\pi^2 M^2 c^4
A_0^4\tau_{*}T}{\hbar^2}.
\end{equation}

This result based on the present 3-dimensional analysis of the
conformal fluctuations can be compared to the analogue 1-dimensional
result that PP found in \cite{powerpercival00}. Using a gaussian
correlation function from the outset they found
\begin{equation*}
\abs{\frac{\delta\rho}{\rho_0}}_{1D}= \frac{\sqrt{\pi} M^2 c^4 A_0^4
\tau_g\,T}{\sqrt{2}\hbar^2},
\end{equation*}
where $\tau_g$ stands for some characteristic correlation time of
the fluctuations. Identifying approximately $\tau_*\simeq\tau_g$, we
have $(32\cII\pi^2)/(\sqrt{\pi/2})\simeq 250$, assuming $\cII \sim
1$. \emph{Thus the present 3-dimensional analysis is seen to predict
a dephasing rate 2 orders of magnitude larger than in the idealized
1-dimensional case.}

\subsection{A remark on the validity of the Dyson expansion}\label{drift}

We have found that the change in the density matrix is given by:
\begin{equation*}
T_2[A_0^4] \sim \left(\frac{M c^2}{\hbar}\right)^2 \tau_{*} T \times
A_0^4,
\end{equation*}
In order for the expansion scheme to be effective, the propagation
time $T$ must be short enough to guarantee that $T_2[A_0^4]$ is
small. How short depends of course on the statistical properties of
the fluctuations, encoded in $\tau_*$, and on the probing particle
mass $M$. A fullerene $C_{70}$ molecule with ($M_{C_{70}} \approx
10^{-24}$ kg) gives $M c^2 / \hbar \approx 10^{27}\mathrm{ s}^{-1}.$
Therefore the approach is consistent only if correlation time
$\tau_{*}$, the flight time $T$ and field squared amplitude $A_0^2$
are appropriately small. We will come back on this issue in
\sref{longdrift}, where it is shown that, in the case of vacuum
fluctuations (introduced in the next session), it is $\tau_{*} \sim
\lambda \Tp$ and $A_0^2 \sim 1/\lambda^2.$ For a flight time $T
\approx 1$ ms, typical of interferometry experiments, this results
in $T_2[A_0^4] \approx 10^7 / \lambda^{-3}$. For any reasonable
value of $\lambda \gtrsim 10^3$ the density matrix change is indeed
small and the Dyson expansion scheme well posed up to second order.
In \ref{ap5} we estimate the fourth order term in the expansion,
which will also yield a term proportional to $A_0^4$. It will be
shown that its contribution in fact vanishes under quite general
circumstances. This puts the result \eref{afe2BBIISS} on an even
stronger basis.

\section{Explicit dephasing rate in the case of vacuum fluctuations}\label{pcf}

The result \eref{afe2BBIISS} is quite general. The only ingredients
entering the analysis so far have been: (i) a spacetime metric
$g_{ab} = (1+A)^2 \eta_{ab}$ with $A = O(\varepsilon \ll 1)$, (ii) a
randomly fluctuating conformal field $A$ satisfying the wave
equation $\pd^c\pd_c A = 0$, and (iii) the isotropic fluctuations
characterized by an arbitrary power spectral density $S(\omega)$.
The dephasing then occurs as a result of the nonlinearity in the
effective potential $V = Mc^2 [\cI A + \cII A^2]$.

A particularly interesting case, potentially related to the
possibility of detecting experimental signs of quantum gravity, is
that in which the fluctuations in $A$ are the manifestation, at the
appropriate semiclassical scale $\ell = \lambda \Lp$, of underlying
\emph{vacuum quantum fluctuations} close to the Planck scale.
Strictly speaking the presence of the probing particle perturbs the
genuine quantum vacuum state. For this reason it would be
appropriate to talk of \emph{effective vacuum}, i.e. up to the
presence of the test particle. By its nature, the present
semiclassical analysis \emph{cannot} take in account the
backreaction of the system on the environment. Therefore we simply
assume that the modifications on the vacuum state can be neglected
as long as the probing particle mass is not too large and the
evolution time short. We thus model the effective vacuum properties
of the conformal field $A$ at the semiclassical scale on the basis
of the properties that real vacuum is expected to possess at the
same scale. It is a fact that vacuum looks the same to all inertial
observers far from gravitational fields. In particular, its energy
density content should be Lorentz invariant. This can obtained
through an appropriate choice of the power spectrum $S(\omega)$.

\subsection{Isotropic power spectrum for `vacuum' conformal fluctuations}

According to the above discussion we expect the average properties
of $A$ above the scale $\ell$ to be Lorentz invariant. In
particular, the interesting quantities derived in \sref{A04s}
\begin{equation*}
   \m{A^2} = \frac{1}{(2\pi)^3}\int \d^3 k\, S(k),
\end{equation*}
\begin{equation*}
   \m{\abs{\nabla A}^2} = \frac{1}{(2\pi)^3}\int \d^3 k\, k^2 S(k),
\end{equation*}
\begin{equation*}
   \m{(\pd_t A)^2} = \frac{1}{(2\pi)^3}\int \d^3 k\, \omega^2(k)
   S(k),
\end{equation*}
should be invariant. As discussed in \ref{ap3}, for a stationary,
isotropic signal, the averages $\m{\cdot}$ can in fact be carried
out through suitable spacetime integrations over an appropriate
averaging scale $L \gg \ell$. In alternative they can be expressed
as in the above integrals depending on the power spectrum and
adopting a high energy cutoff set by $\kcut:=2\pi / (\lambda L_P)$.

The problem of the Lorentz invariance of the above quantities has
been discussed in details by Boyer \cite{boyer1969} within his
random electrodynamics framework. He showed that the choice
\begin{equation}\label{ps}
    S(k):= \frac{\hbar G}{2 c^2}\frac{1}{\omega(k)},
\end{equation}
guarantees a Lorentz invariant measure $\d^3 k/\omega(k)$ (see also
\cite{ryder}) and implies an {energy spectrum} $\varrho(\omega)
\propto \omega^3,$ also shown to be the only possible choice for a
Lorentz invariant energy spectrum of a massless field. The
combination of the constants $\hbar$, $G$ and $c$ gives the correct
dimensions for a power spectrum (i.e. $L^3$), while the factor $1/2$
guarantees that the resulting energy density is equivalent to that
resulting from the superposition of zero-point contributions $\hbar
\omega / 2$. A final important point, which should not be
overlooked, is that Lorentz invariance is preserved provided the
cutoff $\kcut$ is given by the \emph{same number} for all inertial
observers, as also discussed in details by Boyer. In other words
this means that the critical length that sets the border line
between the random field approach and the full quantum gravity
regime is supposed to be the same for \emph{any} inertial observer.
It represents some kind of structural property of spacetime and
\emph{not} an observer dependent property. Accordingly it must not
be transformed under Lorentz transformations. It is important to
note that this requirement will also be satisfied when we employ an
effective cutoff set by the particle Compton length.

\begin{figure}[!b]
\caption{\footnotesize{\textsl{Plot of $R^2(t - t')$. The
adimensional variable $\sigma$ is basically $t - t'$ in units of the
correlation time $\tau_*$. It is seen that $\tau \approx \tau_*$
corresponds to the first of the secondary peaks.}}}\vspace{0.3cm}
\centerline{\psfig{figure=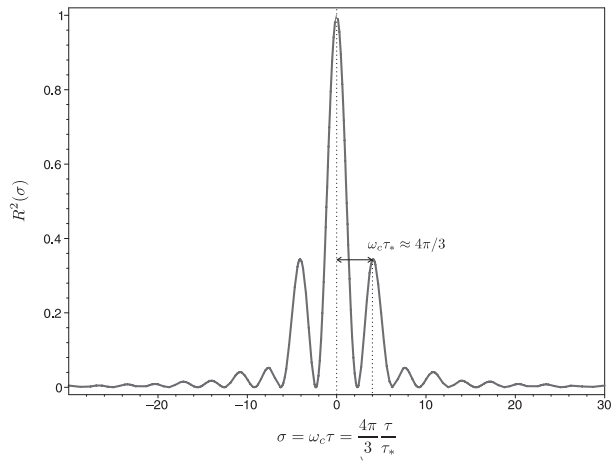,angle=0,width=8cm}} \label{fig2}
\end{figure}
Using \eref{ps} the normalized correlation function can be found
explicitly from the generalized Wiener-Khintchine to be
\begin{equation}\label{normcor}
R(\tau) := \frac{C(\tau)}{C_0} = 2\left[ \frac{\sin\omega_c
\tau}{\omega_c \tau} + \frac{\cos\omega_c \tau -1}{\omega_c^2
\tau^2} \right].
\end{equation}
The peak of the autocorrelation function is linked to the
fluctuations squared amplitude and gives explicitly:
\begin{equation}\label{ap_C20b}
C(0) \equiv A_0^2 = \frac{1}{8\pi\lambda^2},
\end{equation}
implying $\m{A^2} = 1 / (2\lambda)^2$. The correlation time and
characteristic function follow from equations \eref{tts} and
\eref{gamma} as:
\begin{equation}\label{P2_cti}
    \tau_* = \frac{2}{3}\lambda \Tp,
\end{equation}
and
\begin{equation}\label{Gamma}
\Gamma(\sigma) =\frac{3\sin(\sigma)}{\sigma} +
\frac{6\cos(\sigma)}{\sigma^2} -\frac{6\sin(\sigma)}{\sigma^3},
\end{equation}
where $\sigma = \omega_c t$ is a dimensionless variable. The plot of
the squared normalized correlation function $R^2(t-t')$ is shown in
\fref{fig2}: $t-t' = \tau_*$ corresponds to the first secondary peak
in the curve, where the correlation in the fluctuations is reduced
of $\sim 70\%$. This fully motivates the choice of $\tau_*$ to
represent the correlation time.

The explicit form of the characteristic function can be used in
\eref{fff} to evaluate the remaining angular integrals and find the
detailed expression for the density matrix evolution valid for all
$(\xv,\xv')$. Isotropy implies that the result must depend on
$\abs{\xv - \xv'}$ only. The integration is straightforward and
yields the result:
\begin{equation}\label{ffff}
F(\sigma) := 1-\frac{3}{2\sigma^2}\left(1 -
\frac{\sin\sigma\cos\sigma}{\sigma}\right).
\end{equation}
Substituting the results \eref{ap_C20b}, \eref{P2_cti}, \eref{Gamma}
and \eref{ffff} into \eref{afe2BBIISS} yields the explicit result
for the dephasing rate, valid for vacuum fluctuations described by
\begin{figure}[!b]
\caption{\footnotesize{\textsl{Plot of the function $F(\sigma)$ in
the range $\sigma = 0..10$, where $\sigma = 2\pi\abs{\xv-\xv'} /
(\lambda \Lp).$ The curve tends very rapidly to the limiting value
of 1 and for spacial separations $\abs{\xv-\xv'}$ which are slightly
larger than $\ell = \lambda\Lp$ the dephasing rate converges rapidly
to its maximum value.}}}\vspace{0.2cm}\label{fig1}
\centerline{\psfig{figure=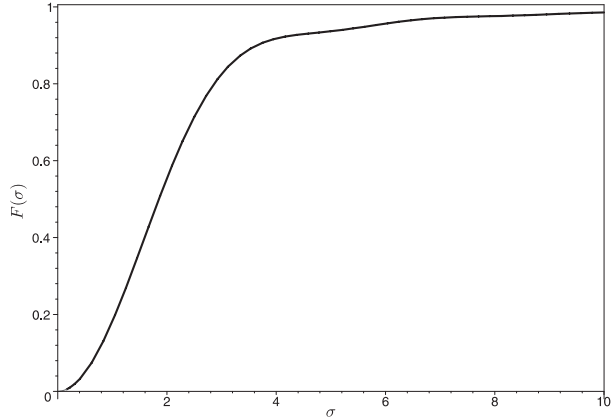,angle=0,width=8cm}}
\end{figure}
$S\sim 1/\omega$:
\begin{equation}\label{fedf2b}
\left|\frac{\delta\rho_{\xv\xv'}}{\rho_0}\right|=
\frac{1}{3\lambda^3}\left(\frac{M}{M_P}\right)^2\left(\frac{T}{T_P}\right)\times
F\left( \frac{2\pi\abs{\xv-\xv'}}{\ell} \right),
\end{equation}
where we considered $\cII \sim 1$ and where
\begin{equation*}
\Mp := \frac{\hbar}{c^2\Tp} = \sqrt{\frac{\hbar c}{G}} = 2.176
\times 10^{-8} \mathrm{kg} = 1.310 \times 10^{19} \mathrm{amu}
\end{equation*}
is the Planck mass. The function $F$ is plotted in \fref{fig1}. It
enjoys the properties $F(0) = 0$ and $F(\sigma) \rightarrow 1$ for
$\sigma \gg 1$, so that for $\abs{\xv-\xv'} \gtrsim 10\ell$ the
decoherence rate converges rapidly to its maximum value.

\section{Discussion}\label{disc}

\subsection{Probing particle resolution scale and effective dephasing
rate}\label{pprs}

Equation \eref{fedf2b} is an important result. It gives the
dephasing rate in the density matrix of a quantum particle
propagating in space under the only action of a randomly fluctuating
potential due to spacetime vacuum conformal fluctuations. The fact
that it predicts an exponential decay of the off diagonal elements
of the system density matrix (which is the distinctive feature of
genuine quantum decoherence) is interesting as a further
confirmation that certain effects involving quantum fluctuations can
be mimicked by means of a semi-classical treatment in the spirit of
Boyer \cite{boyer1969,boyer75}.

A significant feature of our dephasing formula is the quadratic
dependence on the probing particle mass $M$, which comes as a
consequence of the underlying non linearity. The coefficient
$1/(3\lambda^3)$ sets the overall strength of the effect. It is
proportional to $A_0^4$ and to the fluctuations correlation time
$\tau_*$: the more intense the fluctuations, the larger the
dephasing and the longer the various directional component stay
correlated, the higher their ability to induce dephasing. We have
found $\tau_* \approx \lambda \Tp,$ in such a way that the
correlation time directly depends on the spacetime intrinsic cutoff
parameter $\lambda$. According to this picture \emph{all} the
wavelength down to the cutoff $\ell = \lambda \Lp$ should be able to
affect the probing particle. However an atom or molecule is likely
to possess its own resolution scale $\lr$. Thus, whenever $\tau_* c
< \lr$, the ability of the fluctuations to affect the particle would
be reduced, as they would effectively average out. To characterize
this feature of the problem we write, in analogy to $\ell = \lambda
\Lp$,
\begin{equation*}
    \lr = \llr \Lp,
\end{equation*}
and use $\llr$ as a new, \emph{particle dependent}, cutoff
parameter. In general it is $\llr > \lambda$. The new effective
correlation time is now given by $\tau_* \approx \llr \Tp$. The
distance traveled by the fluctuations during a correlation time is
$L_* = c \tau_* \equiv 2\lr / 3$. Thus the effective
\emph{correlation distance} $L_*$ basically corresponds to the
particle resolution scale: short wavelengths that do not keep their
correlation up up the scale $\lr$ average out and cannot affect the
probing particle. The new, effective dephasing rate results by
substituting $\lambda$ with $\llr$ in \eref{fedf2b}:
\begin{equation*}
\left|\frac{\delta\rho_{\xv\xv'}}{\rho_0}\right|=
\frac{1}{3}\left(\frac{\Lp}{\lr}\right)^3\left(\frac{M}{M_P}\right)^2\left(\frac{T}{T_P}\right)\times
F\left( \frac{2\pi\abs{\xv-\xv'}}{\lr} \right).
\end{equation*}

\subsection{Validity of the long drift time regime}\label{longdrift}

\begin{figure}[!t]
\caption{\footnotesize{\textsl{Fourier transform of $R^2(\sigma)$ as
a function of the frequency $\omega$ in units of the cutoff
frequency $\omega_c$.}}}\vspace{0.2cm}
\centerline{\psfig{figure=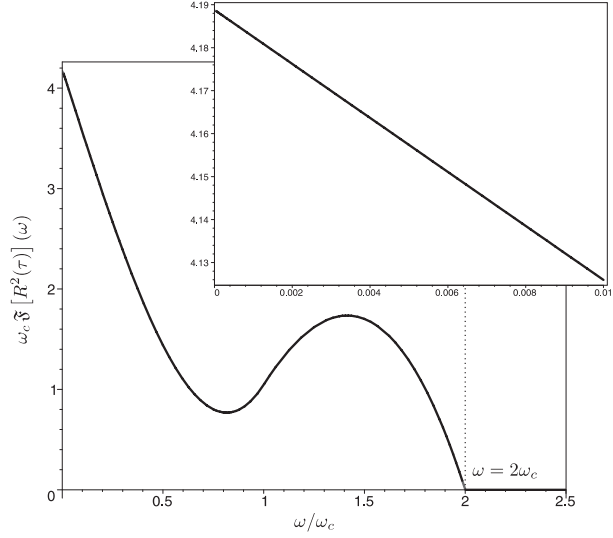,angle=0,width=8cm}} \label{fig3}
\end{figure}

We recall that this results holds for `long' drift times $T$, i.e.
when $T \gtrsim (\Delta \omega)^{-1}$, where $\Delta \omega$ is an
appropriate frequency range over which the Fourier transforms of
$R^2(t)$ and $R(t+\tau)R(t+\tau')$ vary little. We are now in the
position to make this precise and define clearly the limits of
applicability of the theory. To this end we consider the Fourier
transform of $R^2(\omega_c \tau)$:
\begin{equation*}
\mathfrak{F}[R^2(\omega_c \tau)](\omega) =
\frac{1}{\omega_c}\mathfrak{F}[R^2(\sigma)](\omega/\omega_c),
\end{equation*}
with $R(\sigma)$ given in \eref{normcor}. Its plot is displayed in
\fref{fig3}.
The spectrum falls to 0 for $\omega \geq 2\omega_c$. The value of
the peak at $\omega = 0$ is precisely $4\pi / 3$, verifying that
$\tau_* \equiv \mathfrak{F}[R^2(\omega_c \tau)](0) = 4\pi /
(3\omega_c)$. The smaller box shows a zoom of the plot in the region
$\sigma \in [0, 1/100]$: the curve is slow varying in this range
since $\mathfrak{F}[R^2(\omega_c \tau)](0) = 4\pi /3 \approx 4.19$
and $\mathfrak{F}[R^2(\omega_c \tau)](1/100) \approx 4.13$.
Similarly it is possible to check that the Fourier transform of
$R(\sigma + \eta)R( \sigma + \eta')$, where the adimensional
parameters $\eta$ and $\eta'$ depend on space direction and
locations as $\eta:= -\omega_c\kvs\cdot\Delta\xv/c$ and $\eta' :=
-\omega_c\Kvs\cdot\Delta\xv/c$, enjoys a similar property: for
\emph{every} choice of $\eta$ and $\eta'$, the resulting Fourier
transform is slow varying in the range $\sigma \in [0, 1/100]$.
Following this discussion we chose $\Delta\omega \approx [ 0,
\omega_c / 100]$. We can now quantify the concept of `long drift
time' by $T \gtrsim 100 / \omega_c$. From $\omega_c = 2\pi/(\llr
\Tp) = 4\pi/(3\tau_*)$ this yields the condition
\begin{equation*}
T \gtrsim 25 \, \tau_*.
\end{equation*}

\subsection{Some numerical estimates and outlook}\label{sneo}

In summary we have studied the dephasing on a non-relativistic
quantum particle induced by a conformally modulated spacetime
$g_{ab} = (1+A)^2 \eta_{ab}$, where $A$ is a random scalar field
satisfying the wave equation. The important case of vacuum
fluctuations can be characterized by a suitable power spectrum $S
\sim 1/\omega$. If $\lr = \llr \Lp$ is the probing particle
resolution scale, the dephasing rate for $\abs{\xv - \xv'} \gg \lr$
converges rapidly to:
\begin{equation}\label{dododo}
\left|\frac{\delta\rho}{\rho_0}\right|=
\frac{1}{3}\left(\frac{\Lp}{\lr}\right)^3\left(\frac{M}{M_P}\right)^2\left(\frac{T}{T_P}\right).
\end{equation}
The effective correlation time of the fluctuations is given by
$\tau_* \approx \llr \Tp$. The above result holds for `long' drift
times satisfying
\begin{equation*}
T \gtrsim (10 - 10^2) \llr \Tp.
\end{equation*}

\begin{figure}[!b]
\caption{\footnotesize{\textsl{Adimensional cutoff parameter as a
function of the dephasing rate for a $C_{70}$ molecule with a drift
time of $10^{-3}$ s.}}}\vspace{0.2cm}
\centerline{\psfig{figure=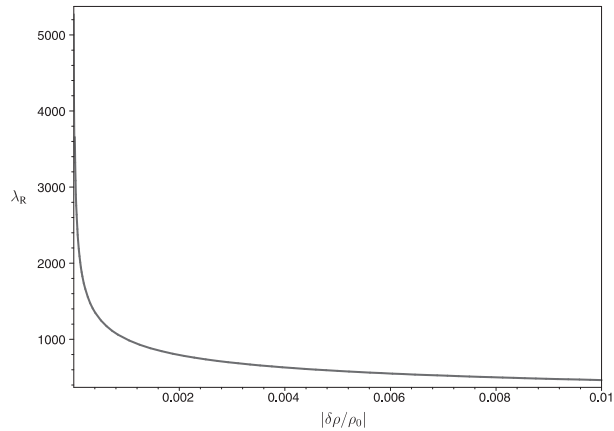,angle=0,width=8cm}} \label{fig4}
\end{figure}

To conclude we want to give some numerical estimates of the
dephasing that could be expected in a typical matter wave
interferometry experiment, e.g. like those described in
\cite{hornberger04}, where fullerene molecules have been employed
with drift times of the order of $T = : \tex \approx 10^{-3}$ s.
Consider e.g. a $C_{70}$ molecule with $M = \mc70 \approx 1.24
\times 10^{-24}$ kg. In comparison to the Planck units we have:
\begin{equation*}
\tex \approx 10^{40}\Tp,\quad \mc70 \approx 10^{-17}\Mp.
\end{equation*}
Thus, it is clear that the most critical factor controlling the
strength of the effect is set by the probing particle mass, together
with the effective resolution cutoff scale. Using these data in
\eref{dododo} we can estimate:
\begin{equation*}
\left|\frac{\delta\rho}{\rho_0}\right| \approx \frac{10^6}{\llr^3}
\quad \Leftrightarrow \quad \llr \approx \left(
\frac{10^6}{\abs{\delta\rho/\rho_0}}
    \right)^{\frac{1}{3}}.
\end{equation*}
\emph{This could be used to estimate $\llr$ if we were able to
identify within an experiment a residual amount of dephasing that
cannot be explained by other standard mechanisms} (e.g.
environmental decoherence, internal degrees of freedom). Figure
\ref{fig4} plots
$\llr$ against $\abs{\delta\rho/\rho_0}$: a dephasing rate due to
conformal fluctuations in the range $1\% - 0.1\%$ would imply a
resolution parameter in the range $\llr \approx 10^3 - 10^4.$ This
would represent a lower bound on $\llr$, as interferometry
experiments will get more and more precise in measuring and modeling
environmental decoherence. Estimating the present typical
uncertainty of typical interferometry experiments as $\abs{\delta
\rho / \rho_0} \approx 0.01\%$ we get
\begin{equation*}
    \llr \gtrsim 10^3,\quad\mathrm{for}\quad C_{70}.
\end{equation*}
We remark that such order of magnitudes estimates are consistent
with a small change in the density matrix and the second order Dyson
expansion approach.

A value for $\llr$ as small as $10^3$ would probably approach the
intrinsic spacetime structural limit set by $\lambda$, i.e. $\ell=
\lambda \Lp$. It is interesting to ask what amount of dephasing our
model predicts, \emph{independently} of experimental data. To this
end we need to prescribe theoretically the particle resolution scale
$\lr$. Though no obvious choice exists, an interesting possibility
would be to set it equal to the particle \emph{Compton length}, i.e.
\begin{equation*}
\lr = \frac{h}{Mc}.
\end{equation*}
This choice is obviously Lorentz invariant and also motivated by the
fact that the Compton length represents a fundamental uncertainty in
the position of a nonrelativistic quantum particle. Indeed, by
Heisenberg uncertainty principle, $\Delta x \approx h/Mc$ would
imply $\Delta p \gtrsim Mc,$ implying an uncertainty in the energy
of the same order of the rest mass $Mc^2$. In such a situation QFT
would become relevant. Alternatively it can also be argued that
wavelengths shorter than $h / Mc$ would have enough energy to create
a particle of mass $M$ from the vacuum. With this choice, equation
\eref{dododo} becomes
\begin{equation}\label{dododo2}
\left|\frac{\delta\rho}{\rho_0}\right|=
\frac{1}{24\pi^3}\left(\frac{M}{M_P}\right)^5\left(\frac{T}{T_P}\right).
\end{equation}
This can be used to predict the amount of dephasing induced by
vacuum conformal fluctuations.

In the case of $C_{70}$ the Compton wavelength is $ \approx
10^{-18}\mathrm{ m}\approx 10^{18}\Lp,$ corresponding to $\llr
\approx 10^{18}$. For a propagation time of $\approx 1$ ms this
gives a dephasing rate
\begin{equation*}
\left|\frac{\delta\rho}{\rho_0}\right|(\mc70,1\mathrm{ ms}) \approx
10^{-44},
\end{equation*}
which would be negligible and far beyond the possibility of
experimental detection. Thus, \emph{in order to achieve dephasing
rate within the current experimental accuracy, much heavier quantum
particles are needed}. In atomic mass units $C_{70}$ has a mass
$\mc70 \approx 10^3$ amu. Equation \eref{dododo2} applied to a
particle with mass $M \approx 10^{11}$ amu and with a drift time $T
\approx 100$ ms gives the estimate:
\begin{equation*}
\left|\frac{\delta\rho}{\rho_0}\right|(10^{11}\mathrm{
amu},100\mathrm{ ms}) \approx 10^{-2}.
\end{equation*}
A drift time of $\sim 100$ ms could possibly be achieved in a space
based experiment. On the other hand, the need of a quantum particle
as heavy as $10^{11}$ amu poses an extraordinary challenge. A
possibility would be to employ quantum entangles states. This is
already being considered in the literature, e.g. in
\cite{everitt08}, where entangled atomic states are studied and
suggested as a possible improved probe for future detection of
spacetime induced dephasing.

The last important point that needs verification is that the
condition \eref{C2} gave earlier at the beginning of this paper is
indeed verified: that was required in order for the change in
momentum due to the fluctuations to be smaller than the `main'
particle momentum $\pl = M\vl$. It read: $ \varepsilon^2 \sim
\m{A^2} \ll (\vl / c)^2$. The field effective mean quadratic
amplitude interacting with the particle is given by $\m{A^2} \sim
{\llr}^{-2}$. Thus we have the condition:
\begin{equation*}
    \frac{1}{\llr} \equiv \frac{\Lp}{\lr} \ll \frac{\vl}{c}.
\end{equation*}
By using the expression for the Planck length and with $\lr$ given
by the particle Compton length, this yields a condition on the
particle mass $M$:
\begin{equation*}
\frac{M}{\Mp} \ll 2\pi\frac{\vl}{c}.
\end{equation*}
For typical laboratory velocities $\vl / c \approx 10^{-6}$ and,
since $\Mp \approx 10^{19}$ amu, this condition is met for particle
masses up to $M \approx 10^{13}$ amu, including the case of $C_{70}$
molecules or the heavier entangled quantum states discussed above.
This limit would be reduced for slower particles.

We conclude by remarking that the theory described in this paper is
quite general, in the sense that as a starting input it only needs a
conformally modulated metric and a scalar field satisfying the wave
equation. Of course, it is important to identify in concrete which
theories of gravity can actually yield such a scenario. This
important problem will be the object of future reports, currently in
preparation: in \cite{bonifacio2009II} a general approach for the
study of fluctuating fields close to the Planck scale is introduced
and the resulting framework applied to standard GR; while
\cite{bonifacio2009III} will consider more general scenarios
involving scalar-tensor theories of gravity.

\begin{ack}
PB wishes to thank L. Galgani, A. Carati, A. Diaferio and D.
Bertacca for the most valuable discussions. Special thanks also go
to C. S. Rodrigues, R. Reis, D. Rittich, the University of Aberdeen
for a Sixth Century Ph.D. Studentship Award and the STFC Centre for
Fundamental Physics.
\end{ack}

\appendix

\section{Stochastic scalar waves and generalized Wiener-Khintchine
theorem}\label{ap3}

\subsection{General solution to the wave equation}

In this appendix we work out some general results holding for scalar
stochastic waves. Working in units with $c=1$, the solution to the
wave equation $(-\partial_t^2 + \nabla^2)f(\v{x},t)=0$ can be
written as
\begin{equation*}
f(\v{x},t) = \frac1{(2\pi)^n} \int \d^n k \tilde{f}(\v{k},t) e^{i
\v{k}\cdot\v{x}},
\end{equation*}
where
\begin{equation*}
\tilde{f}(\v{k},t) = \int \d^n x f(\v{x},t) e^{-i \v{k}\cdot\v{x}}
\end{equation*}
and where $f \in L^2(\mathbb{R}^3)$. The Fourier coefficients take
the general form
\begin{equation*}
\tilde{f}(\v{k},t) = a(\v{k}) e^{-i kt} + b(\v{k}) e^{i kt}
\end{equation*}
for some \emph{complex functions} $a(\v{k})$ and $b(\v{k})$ and
where $k := \abs{\v{k}}$.

The wave can be written as
\begin{equation*}
    \fl f(\v{x},t) = \frac1{(2\pi)^n} \int \d \hat{\v{k}}\,\int_0^{\infty}\!\!\!\d k\, k^{n-1}
    \left[a(\v{k})e^{ik
(\hat{\v{k}}\cdot\v{x}-t)}+b(-\v{k})e^{-ik
(\hat{\v{k}}\cdot\v{x}-t)}\right],
\end{equation*}
where $\d \hat{\v{k}}$ represent the elementary solid angle in
momentum space. Using spherical coordinates in momentum space with
$\v{k}(\vartheta,\varphi,k)$ and employing the notation
$a_{\hat{\v{k}}}(k):= a(\vartheta,\varphi;k)$ and
$b_{-\hat{\v{k}}}(k):= b(\pi-\vartheta,\varphi+\pi;k),$ we define
the \emph{directional wave} along $\hat{\v{k}}(\vartheta,\varphi)$
as
\begin{equation*}
    \fl f_{\hat{\v{k}}}(\hat{\v{k}}\cdot\v{x}/c-t):=\frac1{(2\pi)^n}\int_0^{\infty}\!\!\!\d k\, k^{n-1}
    \left[a_{\hat{\v{k}}}(k)e^{ikc
(\hat{\v{k}}\cdot\v{x}/c-t)}+b_{-\hat{\v{k}}}(k)e^{-ikc(\hat{\v{k}}\cdot\v{x}/c-t)}\right],
\end{equation*}
where for clarity we have restored the speed of light. The general
solution can thus be written according to the directional
decomposition
\begin{equation}\label{ap_Aerdd}
f(\v{x},t) = \int
\d\hat{\v{k}}\,f_{\hat{\v{k}}}(\hat{\v{k}}\cdot\v{x}/c-t).
\end{equation}

\subsection{Stochastic waves}\label{ap_Asw}

In this section we review some elementary properties of stochastic
signals of one time variable $t$.

The fluctuating conformal field at a given space location $\xv$
represents an example of a \emph{stochastic process} $f(t)$. Its
properties can be defined in a statistical sense. If the possible
values of $f$ obey a probability density $p_t(f)$ the statistical
average at time $t$ is defined as
\begin{equation*}
\m{f(t)}:=\int p_t(f)f df.
\end{equation*}
The \emph{variance} is
\begin{equation*}
\m{ [f(t)]^2 } := \int p_t(f)f^2 df.
\end{equation*}
Denoting the probability density of having the particular outcomes
$f(t)$ \emph{and} $f(t')$ by $p_{t t'}(f,f')$, then the
\emph{autocorrelation} or \emph{2-points correlation function} of
$f(t)$ is defined as
\begin{equation*}
R(t-t'):=\m{ f(t)f(t') }:=\int p_{t t'}(f,f') f f' df df'.
\end{equation*}
If the probability of having the value $f(t')$ is completely
independent from the previous outcome $f(t)$ then $p_{t
t'}(f,f')=p_t(f)p_{t'}(f')$ and the stochastic process described by
$f(t)$ is said to be perfectly \emph{uncorrelated}. In this case
$\m{f(t)f(t')}=\m{ f(t) }\m{ f(t') }$. Higher order autocorrelation
functions can also be defined. The \emph{second order correlation
function} is given by
\begin{equation*}
R''(t-t')=\m{ [f(t)]^2[f(t')]^2 } :=\int p_{t t'}(f,f') f^2 f'^2 df
df'.
\end{equation*}

In the case of conformal fluctuations, we assume that the stochastic
process is \emph{stationary}, i.e. all average properties do not
depend on time, and isotropic. This implies that different
directional components have the same statistical properties.
Finally, by assuming the stochastic process to be \emph{ergodic},
and since it satisfies the wave equation, then statistical averages
can be replaced by time or space averages taken on any given sample
function representing the process.

\subsection{Generalized Wiener-Khintchine theorem for a stochastic process
satisfying the wave equation}

We now derive a generalization of the Wiener-Khintchine theorem
linking the power spectrum to the autocorrelation function. We
assume that a particular sample function representing the stochastic
process can be written as in \eref{ap_Aerdd}, in such a way that the
wave equation is satisfied. Some care must be taken in using the
Fourier expansions relations of the previous section. Indeed, for a
given $t$, $f$ \emph{is not} in general a square integrable function
belonging to $L^2(\mathbb{R}^3)$. To circumvent this problem, given
a sample function of the process and for any given $t$, we define
\begin{equation*}
\tilde{f}^L(\v{k},t) := \int_{\mathcal{D}_L} \d^n x f(\v{x},t) e^{-i
\v{k}\cdot\v{x}},
\end{equation*}
where $\mathcal{D}_L$ is a $3$-dimensional cubic domain of side $L$.
Then, the function
\begin{equation}\label{ap_AfeB} f^L(\v{x},t) :=
\frac1{(2\pi)^n} \int \d^n k \tilde{f}^L(\v{k},t) e^{i
\v{k}\cdot\v{x}},
\end{equation}
satisfies the wave equation provided that
\begin{equation}\label{ap_AtilfkB}
\tilde{f}^L(\v{k},t) = a^L(\v{k}) e^{-i kt} + b^L(\v{k}) e^{i kt}.
\end{equation}
These expressions will be used later while taking the limit with $L
\rightarrow \infty$.

Consider now a complex, ergodic, time-dependent stochastic function
$f(\v{x},t)$ in an $n$-dimensional space with time $t$, satisfying
the wave equation $(\partial_t^2 - \nabla^2)f(\v{x},t)=0$. The
autocorrelation function of $f(\v{x},t)$ for any two events
$(\v{x}_1,t_1)$ and $(\v{x}_2,t_2)=(\v{x}_1+\vg{\xi},t_1+\tau)$ is a
function of $\vg{\xi}$ and $\tau$ given by
\begin{equation*}
C(\vg{\xi},\tau)=\m{ f(\v{x}_1,t_1)^*f(\v{x}_2,t_2) },
\end{equation*}
and having the property $C(-\vg{\xi},-\tau) = C(\vg{\xi},\tau)^*$.
For any fixed choice of $\v{x}_1$ and $t_1$ it also satisfies the
wave equation, i.e. $\left(
\partial_\tau^2 - \nabla_{\vg{\xi}}^2
\right)C(\vg{\xi},\tau)=0.$ Assuming that, for any $\tau$,
$C(\vg{\xi},\tau)$ belongs to $L^2(\mathbb{R}^3)$ we have
\begin{equation}\label{ap_ARwsol}
C(\vg{\xi},\tau) = \frac1{(2\pi)^n} \int \d^n k \left[ \a(\v{k})
e^{-i k\tau} + \b(\v{k}) e^{i k\tau} \right]\, e^{i
\v{k}\cdot\vg{\xi}}.
\end{equation}
Notice that $\a(\v{k})^* = \a(\v{k}),\quad \b(\v{k})^* = \b(\v{k}),$
i.e. both $\a(\v{k})$ and  $\b(\v{k})$ are real. For $\tau = 0$
\eref{ap_ARwsol} reduces to the case of stochastic process of one
$n$-dimensional space variable:
\begin{equation*}
C(\vg{\xi},0) = \frac1{(2\pi)^n} \int \d^n k \left[ \a(\v{k}) +
\b(\v{k}) \right]\, e^{i \v{k}\cdot\vg{\xi}}.
\end{equation*}
The standard W-K theorem then implies a power spectrum:
\begin{equation}\label{ap_AredS}
S(\v{k}) = \a(\v{k}) + \b(\v{k}).
\end{equation}

To determine $\a(\v{k})$ and $\b(\v{k})$ we consider the stochastic
process for some fixed time $t_0$. From equations \eref{ap_AfeB} and
\eref{ap_AtilfkB} we have
\begin{equation}\label{ap_AfeC}
f^L(\v{x},t_0) = \frac1{(2\pi)^n} \int \d^n k \tilde{f}^L(\v{k},t_0)
e^{i \v{k}\cdot\v{x}},
\end{equation}
with
\begin{equation}\label{ap_AtilfkC}
\tilde{f}^L(\v{k},t_0) = a^L(\v{k}) e^{-i kt_0} + b^L(\v{k}) e^{i
kt_0}.
\end{equation}
We are thus dealing with one space variable stochastic process for
which the usual W-K theorem holds. Exploiting the fact that the
stochastic process is stationary we define mean power spectral
density as
\begin{equation*}
S(\kv) := \lim_{L,T \rightarrow \infty} \frac{1}{T}\int_0^T \d
    t_0\,\frac{1}{L^n}\m{\tilde{f}^L(\v{k},t_0)^*
\tilde{f}^L(\v{k},t_0)}.
\end{equation*}
Using equations \eref{ap_AfeC} and \eref{ap_AtilfkC} this reduces to
\begin{eqnarray*}
    \fl S(\kv) = \lim_{L \rightarrow \infty}\frac{1}{L^n}\m{\abs{a^L(\v{k})}^2 +
\abs{b^L(\v{k})}^2}.
\end{eqnarray*}
Comparing with \eref{ap_AredS} we have
\begin{equation*}
    S(\v{k})= \lim_{L\rightarrow\infty}\frac{1}{L^n}\m{ \abs{a^L(\v{k})}^2 + \abs{b^L(\v{k})}^2
    }.
\end{equation*}

\subsubsection{Real stochastic process}

If $f(\v{x},t)$ is real it is easily verified that $S(\v{k})$
\emph{is even}, i.e. $S(\v{k})=S(-\v{k})$ and that
$C(\vg{\xi},\tau)$ \emph{is real}. Moreover the power spectrum is
then given simply by
\begin{equation*}
S(\v{k})= \lim_{L\rightarrow\infty}\frac{2}{L^n}\m{
\abs{a^L(\v{k})}^2}.
\end{equation*}
In this case  we get the generalized Wiener-Khintchine theorem for a
real, stationary stochastic scalar in the form:
\begin{equation*}
C(\vg{\xi},\tau)=\frac1{(2\pi)^n} \int \d^n k\, S(\v{k})
\cos(\v{k}\cdot\vg{\xi}- kc\tau),
\end{equation*}
where we have restored the speed of light.

\subsection{Correlation properties of wave components in different
directions}

The results that we have come to establish allow to show that wave
components traveling in different directions are uncorrelated.
Evaluating $\m{ f^*(\v{x}_1,t_1)f(\v{x}_2,t_2) }$ using equation
\eref{ap_Aerdd} we have
\begin{eqnarray}\label{ap_Adc1}
\fl C(\vg{\xi},\tau)=\m{ f^*(\v{x},t)f(\v{x}+\vg{\xi},t+\tau)}
=\nonumber\\
\hspace{-1cm}\int \d\hat{\v{k}}\,\int \d\hat{\v{k}}'\,\m{
f^{*}_{\hat{\v{k}}}(\hat{\v{k}}\cdot\v{x}/c-t)\,
f_{\hat{\v{k}}'}([\hat{\v{k}}'\cdot\v{x}/c-t]+[\hat{\v{k}}'\cdot\vg{\xi}/c-\tau])
}.
\end{eqnarray}
Using the Wiener-Khintchine theorem in its form as given by
\eref{ap_ARwsol}, restoring the speed of light, swapping $\v{k}$ to
$-\v{k}$ in the second integral and splitting the $\d^n k$ integral
in its angular and magnitude parts we can write as well
\begin{equation}\label{ap_Adircor}
    C(\vg{\xi},\tau)=\int
    \d\hat{\v{k}}\,C_{\hat{\v{k}}}(\hat{\v{k}}\cdot\vg{\xi}/c-\tau),
\end{equation}
where we defined the correlation function in the direction
$\hat{\v{k}}$ as
\begin{equation*}
    \fl C_{\hat{\v{k}}}(\hat{\v{k}}\cdot\vg{\xi}/c-\tau):=\frac1{(2\pi)^n}\int_0^{\infty}\!\!\!\d k\, k^{n-1}
 [\a_{\hat{\v{k}}}(k) e^{ikc(\hat{\v{k}}\cdot\vg{\xi}/c-\tau)} + \b_{-\hat{\v{k}}}(k)
 e^{-ikc(\hat{\v{k}}\cdot\vg{\xi}/c-\tau)}],
\end{equation*}
with $\a_{\hat{\v{k}}}(k) := \a(\v{k})$ and $\b_{-\hat{\v{k}}}(k) :=
\b(-\v{k})$. The two equations \eref{ap_Adc1} and \eref{ap_Adircor}
must be equivalent. This implies at once the following equation
\begin{equation*}
    \fl \m{ f^{*}_{\hat{\v{k}}}(\hat{\v{k}}\cdot\v{x}/c-t)\,
f_{\hat{\v{k}}'}([\hat{\v{k}}'\cdot\v{x}/c-t]+[\hat{\v{k}}'\cdot\vg{\xi}/c-\tau])
}=
\delta(\v{k},\v{k}')\,C_{\hat{\v{k}}}(\hat{\v{k}}\cdot\vg{\xi}/c-\tau)
\end{equation*}
or, equivalently, since $\hat{\v{k}}\cdot\v{x}/c$ has the dimensions
of a time,
\begin{equation}\label{ap_Afff2}
     \m{ f^{*}_{\hat{\v{k}}}(t)\,
f_{\hat{\v{k}}'}(t+\tau) }=
\delta(\v{k},\v{k}')\,C_{\hat{\v{k}}}(\tau).
\end{equation}
We thus see that the fluctuating field can be resolved into
components along different directions represented by completely
uncorrelated functions of \emph{just one time variable}.

\subsection{Isotropic power spectrum and field averages}\label{A04s}

In the relevant case of a real and isotropic signal in 3-dimensional
space the spectral density must also be isotropic, i.e. $S(\v{k}):=
S(k)$. We can then introduce an \emph{isotropic correlation
function} as
\begin{equation}\label{ap_Adsp2}
    C(\tau) := \frac{1}{(2\pi)^3}\int_0^{\infty}\!\!\!\d k\,
    k^2 S(k)\cos(kc\tau)
\end{equation}
and \eref{ap_Afff2} reads simply $\m{ f_{\hat{\v{k}}}(t)\,
f_{\hat{\v{k}}'}(t+\tau) }= \delta(\v{k},\v{k}')\,C(\tau)$. Using
this together with \eref{ap_Aerdd}, we have
\begin{equation*}
    \fl\m{ f(\xv,t)^2}=\m{\int\d\kvs f_{\kvs}(\xv,t)\int\d\kvs'
    f_{\kvs'}(\xv,t)} =C_0\int\d\kvs
    \int\d\kvs'\delta(\v{k},\v{k}'),
\end{equation*}
so that
\begin{equation*}
    \m{ f(\xv,t)^2}=4\pi C_0.
\end{equation*}
Using this with equation \eref{ap_Adsp2} we have
\begin{equation*}
\m{ f(\xv,t)^2} = \frac1{(2\pi)^3} \int \d^3 k\, S(k).
\end{equation*}
It is also useful to deduce two expressions for the average of the
square of the time and space derivatives of the field, as these are
directly related to the field energy density. Using equation
\eref{ap_AfeB} and the ergodic property it is straightforward to
show that
\begin{equation*}
\m{ \left(\partial_t f\right)^2 } = \frac1{(2\pi)^3} \int \d^3 k\,
k^2 S(\kv),
\end{equation*}
and
\begin{equation*}
\m{ \abs{\nabla f}^2 } = \frac1{(2\pi)^3} \int \d^3 k\, k^2 S(\kv).
\end{equation*}

\section{Treatment of the term $T_4$}\label{apx}

We consider the correction term $T_4 = -i\hbar\left( \dot{A} -
A\dot{A} \right)\psi - i\hbar \vl \left( A_{,x} - A A_{,x}
\right)\psi,$ derived in \sref{lvle}. Writing $A$, in the isotropic
and real case, as:
\begin{equation}\label{sig1}
A \approx \int \d \hat{\v{k}}\,\int_0^{2\pi/\lr}\!\!\!\d k\, k^2
a(k)e^{ik (\hat{\v{k}}\cdot\v{x}-ct)},
\end{equation}
where the upper cutoff is set by the particle resolution scale
$\lr.$ The power spectrum is basically proportional to the square of
the Fourier component $a(k)$. For $S(k) \sim 1 / k$ we have $a(k)
\sim k^{-1/2}$, so that the effective coefficient appearing in the
expansion is $k^2 a(k) \sim k^{3/2}$. Thus the short wavelengths
close to the cutoff give the most important contribution. For this
reason we can approximate the field as
\begin{equation}\label{sig2}
A \approx \int \d \hat{\v{k}}\,\int_0^{2\pi/\lr}\!\!\!\d k\, k^{3/2}
\Delta(k - k_A)e^{ik (\hat{\v{k}}\cdot\v{x}-ct)},
\end{equation}
where the function $\Delta(k - k_A)$ is peaked around a typical wave
number $k_A$. This can in principle be selected in such a way that
the average properties of \eref{sig1} are equivalent to those of
\eref{sig2}. Effectively we get:
\begin{equation}\label{sig3}
A \approx \int \d \hat{\v{k}} {k_A}^{3/2} e^{ik_A
(\hat{\v{k}}\cdot\v{x}-ct)},
\end{equation}
so that the conformal field is approximated as a fast varying and
isotropic random signal characterized by a \emph{single typical
wavelength} $\lambda_A \equiv 2\pi / k_A$. In relation to the
fluctuations ability to affect the particle, this will be close to
the particle resolution scale, i.e. we put $\lambda_A \equiv \kappa
\lr$, with $\kappa \gtrsim 1$. From \eref{sig3} we now have:
\begin{equation*}
\dot{A} \approx -i k_A c A = -\frac{2\pi i}{\kappa\lr}cA = - \frac{i
Mc^2}{\kappa \hbar}A,
\end{equation*}
where we used $\lr = h/Mc$. The space derivatives yield:
\begin{equation*}
A_{,x} \approx  i{k_A}^{5/2} \int \d \hat{\v{k}} \hat{k}_x e^{ik_A
(\hat{\v{k}}\cdot\v{x}-ct)} = 0.
\end{equation*}
Using these two relations in \eref{t4} yields the result
\eref{t4bis}.

\section{An integral identity}\label{ap4}

\begin{quote}
In this appendix we prove that the result
\begin{equation*}
\fl I := \int_{0}^{T}\!\!\!\d t \left[2 \int_{0}^{t}\!\!\d t'
f(t-t')- \int_{0}^{T}\!\!\!\d t'f(t-t'-\kvs\cdot\Delta \xv)\right] =
0
\end{equation*}
holds for an \emph{arbitrary even function} $f$ and in the limit $\kvs\cdot\Delta\xv / T \rightarrow 0$.\\
\end{quote}
For simplicity let $\Delta := \kvs\cdot\Delta \xv.$ Defining the
variable $\tau:= t-t'$ we have
\begin{equation*}
    \int_{0}^{t}\!\!\d t'
f(t-t') = \int_0^t\!\!\!\d\tau f(\tau),
\end{equation*}
while, with $\tau:= t-t'-\Delta$,
\begin{equation*}
    \fl\int_{0}^{T}\!\!\!\d
t'f(t-t'-\Delta)=\int_{t-\Delta - T}^{t-\Delta}\!\!\!\d\tau
f(\tau)=\int_0^{t-\Delta}\!\!\!\d\tau f(\tau) - \int_0^{t-\Delta -
T}\!\!\!\d\tau f(\tau).
\end{equation*}
Introducing the primitive of $f(t)$
\begin{equation*}
F(t):=\int_0^t\!\!\!\d\tau f(\tau)
\end{equation*}
we can re-write $I$ as
\begin{equation*}
I=2\int_{0}^{T}\!\!\!{\d}t F(t)-\int_{0}^{T}\!\!\!{\d}t\,
F(t-\Delta) + \int_{0}^{T}\!\!\!{\d}t\, F(t-\Delta -T).
\end{equation*}
Performing a further change of variable $z:= t - \Delta$ the second
integral reads
\begin{eqnarray*}
\fl \int_{0}^{T}\!\!\!{\d}t\, F(t-\Delta) = \int_{-\Delta}^{T -
\Delta}\!\!\!dz\, F(z) := \int_{0}^{T - \Delta}\!\!\!dz\, F(z) -
\int_{0}^{- \Delta}\!\!\!dz\, F(z).
\end{eqnarray*}
Performing a similar operation on the third integral we obtain
\begin{equation*}
\fl I=2\int_{0}^{T}\!\!\!{\d}t F(t)- \int_{0}^{T - \Delta}\!\!\!dz\,
F(z) + \int_{0}^{- \Delta}\!\!\!dz\, F(z) + \int_{0}^{-
\Delta}\!\!\!dz\, F(z) - \int_{0}^{-T - \Delta}\!\!\!dz\, F(z).
\end{equation*}
Now we use the information $T \gg \Delta$ and $f(t)=f(-t)$. As an
elementary consequence we have that $F(t)$ is $F(t)=-F(-t)$,
implying that
\begin{equation*}
\int_{0}^{\chi}\!\!\!\d\tau F(\tau)=\int_{0}^{-\chi}\!\!\!\d\tau
F(\tau)
\end{equation*}
Approximating $T \pm \Delta \approx T$ and swapping the sign of the
upper integration bound appropriately we have
\begin{eqnarray*}
\fl I=2\int_{0}^{T}\!\!\!{\d}t F(t)- \left[\int_{0}^{T}\!\!\!dz\,
F(z) - \int_{0}^{\Delta}\!\!\!dz\, F(z)\right] - \left[
\int_{0}^{T}\!\!\!dz\, F(z) - \int_{0}^{\Delta}\!\!\!dz\,
F(z)\right].
\end{eqnarray*}
The integrals from 0 to $\Delta$ can all be neglected exploiting
again the fact that $T \gg \Delta$ and we obtain
\begin{eqnarray*}
I \approx 2\int_{0}^{T}\!\!\!{\d}t F(t)-2\int_{0}^{T}\!\!\!{\d}t
F(t)=0.
\end{eqnarray*}
The result in exact in the limit $\Delta / T \rightarrow 0.$


\section{Fourth order term in the Dyson expansion}\label{ap5}

The fourth order propagator in the Dyson expansion is
\begin{equation}
\fl \Ko_4(T):=\left(\frac{-i}{\hbar}\right)^4\!\!
\int_{0}^{T}\!\!{\d}t^{(1)}\!\!\int_{0}^{t^{(1)}}\!\!\!{\d}t^{(2)}\!\!\int_{0}^{t^{(2)}}\!\!\!{\d}t^{(3)}\!\!
\int_{0}^{t^{(3)}}\!\!\!\!{\d}t^{(4)}{\Ho}(t^{(1)}){\Ho}(t^{(2)}){\Ho}(t^{(3)}){\Ho}(t^{(4)}),
\end{equation}
implying a fourth order term in the expression for the density
matrix evolution given by
\begin{small}
\begin{eqnarray*}
\fl\m{\Ko_4 \rho_0} \sim
\int_{0}^{T}\!\!{\d}t^{(1)}\!\!\int_{0}^{t^{(1)}}\!\!\!{\d}t^{(2)}\!\!\int_{0}^{t^{(2)}}\!\!\!{\d}t^{(3)}\!\!
\int_{0}^{t^{(3)}}\!\!\!\!{\d}t^{(4)} \int \d^3 y^{(1)}
\Pr{\yv^{(1)}}\cdots\int \d^3 y^{(4)}\Pr{\yv^{(4)}}\times\\
\times\m{V(\yv^{(1)},t^{(1)})V(\yv^{(2)},t^{(2)})V(\yv^{(3)},t^{(3)})V(\yv^{(4)},t^{(4)})}\rho_0.
\end{eqnarray*}
\end{small}
Since the potential is $V = Mc^2 (\cI A + \cII A^2)$ the average
yields one term proportional to $A_0^4$:
\begin{eqnarray*}
\fl\m{V(\yv^{(1)},t^{(1)})V(\yv^{(2)},t^{(2)})V(\yv^{(3)},t^{(3)})V(\yv^{(4)},t^{(4)})}
\Rightarrow\\
\hspace{-1cm}\left(\frac{Mc^2 A_0}{\hbar}\right)^4\int\!
\d\kvs^{(1)}\cdots\int\!
\d\kvs^{(4)}\m{f_{\kvs}(\tau^{(1)})f_{\kvs'}(\tau^{(2)})f_{\kvs}(\tau^{(3)})f_{\kvs'}(\tau^{(4)})},
\end{eqnarray*}
where $\tau^{(i)} := t^{(i)} - \yv^{(i)}\cdot\kvs^{(i)}$. This
requires knowledge of the 4-point correlation function, involving
the average of the product of 4 directional components evaluated at
different points. For a real random process having a zero mean and
gaussian distribution the 4-points function reduces to
\cite{adler81,erkisen02}:
\begin{eqnarray*}
\fl\m{f_{\kvs}(\tau^{(1)})f_{\kvs'}(\tau^{(2)})f_{\kvs}(\tau^{(3)})f_{\kvs'}(\tau^{(4)})}
=
\m{f_{\kvs}(\tau^{(1)})f_{\kvs'}(\tau^{(2)})}\m{f_{\kvs}(\tau^{(3)})f_{\kvs'}(\tau^{(4)})}\\
+\m{f_{\kvs}(\tau^{(1)})f_{\kvs'}(\tau^{(3)})}\m{f_{\kvs}(\tau^{(2)})f_{\kvs'}(\tau^{(4)})}\\
+\m{f_{\kvs}(\tau^{(1)})f_{\kvs'}(\tau^{(4)})}\m{f_{\kvs}(\tau^{(2)})f_{\kvs'}(\tau^{(3)})}.
\end{eqnarray*}
We can now use equation \eref{e17} to express the 2-point
correlations:
\begin{eqnarray*}
\fl
\m{f_{\kvs}(\tau^{(1)})f_{\kvs'}(\tau^{(2)})f_{\kvs}(\tau^{(3)})f_{\kvs'}(\tau^{(4)})}
=\\
\hspace{0.3cm}\delta(\kvs^{(1)},\kvs^{(2)})R(\tau^{(1)}-\tau^{(2)})\delta(\kvs^{(3)},\kvs^{(4)})R(\tau^{(3)}-\tau^{(4)})\\
+\delta(\kvs^{(1)},\kvs^{(3)})R(\tau^{(1)}-\tau^{(3)})\delta(\kvs^{(2)},\kvs^{(4)})R(\tau^{(2)}-\tau^{(4)})\\
+\delta(\kvs^{(1)},\kvs^{(4)})R(\tau^{(1)}-\tau^{(4)})\delta(\kvs^{(2)},\kvs^{(3)})R(\tau^{(2)}-\tau^{(3)}).
\end{eqnarray*}
This implies that the term $T_4[A_0^4]$ deriving from $\m{\Ko_4
\rho_0}$ has the structure:
\begin{equation*}
\fl T_4[A_0^4] \sim 3 \left[\int\! \d\kvs^{(1)}\!\!\int\!\!
\d\kvs^{(2)}\!\!
\int_{0}^{T}\!\!\!{\d}t^{(1)}\!\!\int_{0}^{T}\!\!\!{\d}t^{(2)}
\delta(\kvs^{(1)},\kvs^{(2)})R(\tau^{(1)}-\tau^{(2)}) \right]^2,
\end{equation*}
where all upper bounds in the time integration can be set equal to
$T$ by appropriate normal ordering \cite{roman65}. By carrying out
one of the two angular integrations we have:
\begin{equation*}
\fl T_4[A_0^4] \sim 3 \left[\int\! \d\kvs^{(1)}\!\!
\int_{0}^{T}\!\!\!{\d}t^{(1)}\!\!\int_{0}^{T}\!\!\!{\d}t^{(2)}
R[t^{(1)} - t^{(2)} - \kvs^{(1)}\cdot(\yv^{(1)}-\yv^{(2)})]
\right]^2.
\end{equation*}
The double time integral can be simplified using the general result
\eref{TTtt01} and we have
\begin{equation*}
T_4[A_0^4] \sim 3 \left[\int\! \d\kvs^{(1)} \mathfrak{F}\left[R(t +
\tau)\right](0) T \right]^2,
\end{equation*}
where $\mathfrak{F}$ denotes Fourier transform and $\tau :=
-\kvs^{(1)}\cdot(\yv^{(1)}-\yv^{(2)}). $ The Fourier transform can
be evaluated using that $ R(t + \tau)= C(t+\tau) / C_0.$ Then
\begin{small}
\begin{eqnarray*}
\fl\mathfrak{F}\left[R(t + \tau)\right](\omega) =
\frac{1}{C_0}\int_{-\infty}^{\infty}\d t C(t+\tau) e^{-i\omega t}
\nonumber\\
= \frac{1}{C_0(2\pi c)^3} \int_0^{\omega_c}\!\!\d
\omega'\int_{-\infty}^{\infty}\d t\, \omega'^2 S(\omega')
e^{-i\omega t} \cos \omega' (t+\tau)\nonumber\\
= \frac{1}{2C_0(2\pi c)^3} \int_0^{\omega_c}\!\!\d
\omega'\int_{-\infty}^{\infty}\d t\, \omega'^2 S(\omega')
e^{-i\omega t} \left[ e^{i\omega'(t+\tau)} + e^{-i\omega'(t+\tau)}\right]\nonumber\\
= \frac{1}{2C_0(2\pi c)^3} \int_0^{\omega_c}\!\!\d \omega' \omega'^2
S(\omega') \int_{-\infty}^{\infty}\d t\, \left[
e^{i(\omega'-\omega)t}e^{i\omega'\tau} +
e^{i(-\omega'-\omega)t}e^{-i\omega'\tau}\right].
\end{eqnarray*}
\end{small}
Integrating with respect to $t$ gives
\begin{eqnarray*}
\fl\mathfrak{F}\left[R(t + \tau)\right](\omega) =\frac{1}{2C_0(2\pi
c)^3} \int_0^{\omega_c}\!\!\d \omega' \omega'^2 S(\omega') \left[
\delta(\omega'-\omega)e^{i\omega'\tau} +
\delta(-\omega'-\omega)e^{-i\omega'\tau}\right].
\end{eqnarray*}
This vanishes for $\omega > \omega_c$. For $\omega = 0$ we have:
\begin{eqnarray*}
\fl\mathfrak{F}\left[R(t + \tau)\right](0) =\frac{1}{2C_0(2\pi c)^3}
\int_0^{\omega_c}\!\!\d \omega' \omega'^2 S(\omega') \left[
\delta(\omega')e^{i\omega'\tau} +
\delta(-\omega')e^{-i\omega'\tau}\right].
\end{eqnarray*}
Carrying out the frequency integral and using the properties of the
$\delta$ function we have:
\begin{eqnarray*}
\mathfrak{F}\left[R(t + \tau)\right](0) &=\frac{1}{C_0(2\pi c)^3}
\lim_{\omega\rightarrow 0} \omega^2 S(\omega).
\end{eqnarray*}
In the interesting case $S \sim 1/\omega$ this tends to 0, in such a
way that $T_4[A_0^4] \rightarrow 0$ and the fourth order Dyson
expansion term doesn't give any contribution.




\end{document}